\documentclass[notitlepage,12pt,groupedaddress,secnumarabic,prd,showpacs,showkeys,onecolumn,nofootinbib]{revtex4}
\usepackage{amsfonts}
\usepackage{amsmath}
\usepackage{amssymb}
\usepackage{graphicx}
\usepackage{appendix}
\usepackage{color}

\begin{document}

\title{Interacting kinks and meson mixing}
\author{J.R. Morris}
\affiliation{Physics Department, Indiana University Northwest, 3400 Broadway, Gary,
Indiana 46408, USA}
\email{jmorris@iun.edu}

\begin{abstract}
A Rayleigh-Schr\"{o}dinger type of perturbation scheme is employed to study
weakly interacting kinks and domain walls formed from two different real
scalar fields $\chi$ and $\varphi$. An interaction potential $%
V_{1}(\chi,\varphi)$ is chosen which vanishes in a vacuum state of either
field. Approximate first order corrections for the fields are found, which
are associated with scalar field condensates inhabiting the zeroth order
topological solitons. The model considered here presents several new and
interesting features. These include (1) a condensate of \textit{each} kink
field inhabits the \textit{other} kink, (2) the condensates contribute an
associated mass to the system which vanishes when the kinks overlap, (3) a
resulting mass defect of the system for small interkink distances allows the
existence of a loosely bound state when the interkink force is repulsive. An
identification of the interaction potential energy and forces allows a
qualitative description of the classical motion of the system, with bound
states, along with scattering states, possible when the interkink force is
attractive. (4) Finally, the interaction potential introduces a mixing and
oscillation of the perturbative $\chi$ and $\varphi$ meson flavor states,
which has effects upon meson-kink interactions.
\end{abstract}

\pacs{11.27.+d, 98.80.Cq}
\keywords{topological soliton, kink interaction, domain walls, meson mixing}
\maketitle

\section{Introduction}

\ \ It has been long recognized that certain nonlinear field theories
possessing multiple disconnected vacuum states admit, in addition to a set
of perturbative particle spectra, additional states associated with the
topology of the vacuum manifold (see, e.g., \cite{Goldstone75},\cite%
{Jackiw77}, and references therein). These nonperturbative states, or
\textquotedblleft solitons\textquotedblright , typically have nontrivial
internal structures that can depend upon one or more spatial dimensions
(see, e.g., \cite{Goldstone75}-\cite{KTbook}). Studies of one dimensional
topological defects, describing \textquotedblleft kinks\textquotedblright ,
or planar domain walls can expose interesting properties of solitons and
their interactions with other solitons and ordinary matter. In addition, the
one dimensional defects can be described by sets of simpler differential
equations that depend only upon one space variable. In addition, much
attention has been given to investigations involving various interactions
between scalar fields describing kinks of more than one variety, which can
arise from models involving two distinct scalar fields. (For a sample of
such types of analyses see, for example, \cite{Baz PLA13}-\cite{Alfonso
PRD07}.)

\bigskip

\ \ Here, a fairly simple model describing two weakly interacting scalar
fields, denoted by $\chi $ and $\varphi $, is presented which exhibits some
new and interesting features. A potential $V(\chi ,\varphi )=V_{0}(\chi
,\varphi )+V_{1}(\chi ,\varphi )$ is chosen which admits solutions
describing interacting $\phi ^{4}$-type kink solutions in a $1+1$
dimensional spacetime (or planar domain walls in higher dimensional
spacetimes. These are simply referred to here as \textquotedblleft
kinks\textquotedblright\ for simplicity (see, e.g., \cite{Goldstone75}-\cite%
{KTbook})). The unperturbed potential is given by $V_{0}(\chi ,\varphi )=%
\frac{1}{4}\lambda _{\chi }\left( \chi ^{2}-\eta ^{2}\right) ^{2}+\frac{1}{4}%
\lambda _{\varphi }\left( \varphi ^{2}-\sigma ^{2}\right) ^{2}$ and the
interaction potential is chosen to be $V_{1}(\chi ,\varphi )=\frac{1}{2}%
\beta \left( \chi ^{2}-\eta ^{2}\right) \left( \varphi ^{2}-\sigma
^{2}\right) $ where the parameter $\beta $ is small in comparison to $%
\lambda _{\chi }$ or $\lambda _{\varphi }$, i.e., $|\beta |\ll \lambda $.
When $\beta =0$ the model admits the familiar tanh - like solutions
describing $\chi $ and $\varphi $ kinks and antikinks, and when the
interaction is turned on with $\beta \neq 0$ the $\chi $ and $\varphi $
kinks interact with each other. We allow the nonvanishing $\beta $ to be
either positive or negative, allowing for either repulsive or attractive
interactions between the $\chi $ and $\varphi $ kinks. We note that $V_{1}$
vanishes in the vacuum state of either field, where $\chi _{\text{vac}}=\pm
\eta $ and $\varphi _{\text{vac}}=\pm \sigma $, i.e., the vacuum states are
preserved by the interaction.

\bigskip

\ \ The addition of a perturbing potential necessitates corrections to the $%
\tanh kx$ kink solutions of the unperturbed theory. (See, for example, \cite%
{Deform}-\cite{Defect13} and \cite{DB96},\cite{DB97}.) Here, a basic
Rayleigh-Schr\"{o}dinger type of perturbation scheme is developed to obtain
equations describing a set of corrections $\left\{ \chi_{n},\varphi
_{n}\right\} $ to the unperturbed base solutions $\chi_{0},\varphi_{0}$. We
focus upon the first order static corrections $\chi_{1}(x)$ and $\varphi
_{1}(x)$, each of which satisfies a nonhomogeneous linear differential
equation (DE) involving hyperbolic functions. Failing to find exact
analytical solutions to these equations, we instead obtain approximate
analytical solutions using a type of \textquotedblleft thin
wall\textquotedblright\ approximation. These approximate analytical
solutions have the advantage of displaying the roles of the various model
parameters, such as the widths $w_{\chi}$ and $w_{\varphi}$ of the kinks and
the separation distance $a$ between them. These solutions are useful in
obtaining subsequent features of the model. Although the approximation is
expected to work better for massive, narrow width kinks, it is expected to
exhibit, at least, the qualitative behaviors of less massive, wider, kinks
as well.

\bigskip

\ \ The first order corrections for this model yield some surprising
results, some of which may apply to other two-field models, as well. (1) One
surprising result is that the static first order corrections, which are
associated with scalar field condensates, have the peculiar character that
they are pronounced at the locations of the $\chi_{0}$ and $\varphi_{0}$
kinks. More specifically, the condensate of \textit{each} kink field
inhabits the \textit{other} kink. That is, $\chi_{1}(x)$ becomes localized
at the location of the $\varphi_{0}$ kink, and $\varphi_{1}(x)$ is localized
at the location of the $\chi_{0}$ kink. These localized corrections, or
\textquotedblleft displaced scalar field condensates\textquotedblright, are
nontopological and essentially inhabit the zeroth order kinks. The
combination of a soliton with a scalar field condensate then comprises a
\textquotedblleft structured\textquotedblright\ kink.

\bigskip

\ \ (2) The approximate \textquotedblleft mass\textquotedblright\ associated
with each condensate is found, which contributes to the total mass of the
structured kink. A distinctive property of the structured kinks is that the
mass associated with a condensate decreases with separation distance between
the kinks when they are close together, and the condensate mass vanishes
when the kinks overlap, i.e., occupy the same position. (3) The resulting
\textquotedblleft mass defect\textquotedblright, or \textquotedblleft
binding energy\textquotedblright, connected with the condensate masses
therefore allows the existence of a loosely bound state of the $\chi$ and $%
\varphi$ kinks when the interkink force is repulsive.

\bigskip

\ \ Using the base solutions for the static kinks, a classical potential
energy of interaction, along with an interaction force between the kinks,
can be defined, allowing a qualitative description of the classical motion
of the system. The interaction force can be either attractive ($\beta<0$) or
repulsive ($\beta>0$).\ \ When the interkink force is attractive, stronger
bound states can exist, giving rise to composite two-kink states of the $%
(\chi,\varphi)$ system. These composite states can have topological charges
of $Q=\pm2$ or $0$.

\bigskip

\ \ (4) Finally, it is pointed out that an interaction between the $\chi$
and $\varphi$ fields produces a nondiagonal mass matrix for the perturbative
\textquotedblleft meson\textquotedblright\ flavor states $|\chi\rangle$ and $%
|\varphi\rangle$ that are built from the vacuum states $\chi_{\text{vac}%
}=\pm\eta$ and $\varphi_{\text{vac}}=\pm\sigma$. Therefore, the flavor
states $|\chi\rangle$ and $|\varphi\rangle$ are combinations of mass
eigenstates $|\phi_{1}\rangle$ and $|\phi_{2}\rangle$, resulting in
oscillations of the $\chi$ and $\varphi$ scalar particles. Since only $\chi$
($\varphi$) particles reflect from a $\varphi$ ($\chi$) kink, the radiative
force exerted on one kink due to scalar radiation from the other kink will
be affected by the oscillations.

\bigskip

\ \ Computational details for several results are relegated to Appendices.

\section{The Model}

\ \ We take the Lagrangian of real-valued scalar fields $\chi$ and $\varphi$
to be%
\begin{equation}
\mathcal{L}=\frac{1}{2}(\partial\chi)^{2}+\frac{1}{2}(\partial\varphi
)^{2}-V(\chi,\varphi),\ \ \ \ \
V(\chi,\varphi)=V_{0}(\chi,\varphi)+V_{1}(\chi,\varphi)  \label{1}
\end{equation}
with $V_{1}$ acting as a small perturbation to $V_{0}$, where%
\begin{equation}
\begin{array}{ll}
& V_{0}(\chi,\varphi)=\frac{1}{4}\lambda_{\chi}\left( \chi^{2}-\eta
^{2}\right) ^{2}+\frac{1}{4}\lambda_{\varphi}\left( \varphi^{2}-\sigma
^{2}\right) ^{2} \\ 
& V_{1}(\chi,\varphi)=\frac{1}{2}\beta\left( \chi^{2}-\eta^{2}\right) \left(
\varphi^{2}-\sigma^{2}\right)%
\end{array}
\label{2}
\end{equation}

where the coupling constants $\lambda_{\chi}$ and $\lambda_{\varphi}$ are
positive, and $\beta$ can be either positive or negative. When $V_{1}=0$ the
equations of motion support the familiar $\phi^{4}$ type of kink/domain wall
solutions. The $\frac{1}{2}\beta\chi^{2}\varphi^{2}$ term is an interaction
term and $V_{1}$ is considered to be a small perturbation with $|\beta
|\ll\lambda_{\chi},\lambda_{\varphi}$.

\bigskip

\ The equations of motion are given by $\square\chi+\partial_{\chi}V(\chi,%
\varphi)=0$ and$\ \square\varphi+\partial_{\varphi}V(\chi,\varphi)=0$, or,
more specifically, by 
\begin{subequations}
\label{3}
\begin{align}
\square\chi+\lambda_{\chi}\chi(\chi^{2}-\eta^{2})+\beta\chi(\varphi^{2}-%
\sigma^{2}) & =0,  \label{3a} \\
\ \ \
\square\varphi+\lambda_{\varphi}\varphi(\varphi^{2}-\sigma^{2})+\beta%
\varphi(\chi^{2}-\eta^{2}) & =0  \label{3b}
\end{align}

where $\square=\partial_{t}^{2}-\nabla^{2}$ and $\partial_{\chi}=\partial/%
\partial\chi$, $\partial_{\varphi}=\partial/\partial\varphi$, etc. The
vacuum states are $\chi_{\text{vac}}=\pm\eta$ and $\varphi_{\text{vac}%
}=\pm\sigma$.

\bigskip

\ \ In the absence of interaction ($\beta=0$) the system admits static kink
solutions $\chi_{0}(x)=\eta\tanh k_{\chi}(x-x_{\chi})$ and $\varphi
_{0}(x)=\sigma\tanh k_{\varphi}(x-x_{\varphi})$, where the parameter $%
k_{\chi,\varphi}$ is the inverse width of the kink, $k_{\chi,\varphi
}=1/w_{\chi,\varphi}$, and $x_{\chi,\varphi}$ is the position of the kink
center where $\chi_{0}(x)=0$ and $\varphi_{0}(x)=0$. The antikink solutions
are $\bar{\chi}_{0}(x)=-\chi_{0}(x)$ and $\bar{\varphi}_{0}(x)=-\varphi
_{0}(x)$. In addition, there are excitation modes of these kink solutions,
including a continuum of meson states $\chi_{p}(x,t)$ and $\varphi_{p}(x,t)$
for each field with momentum $p$ and particle masses $m_{\chi}$ and $%
m_{\varphi}$. The one dimensional kink solutions (or planar domain wall
solutions) take values $\chi_{0}=\pm\eta$ and $\varphi_{0}=\pm\sigma$
asymptotically. The $\chi$ and $\varphi$ \textquotedblleft
meson\textquotedblright\ (i.e., perturbative) particle masses are given by $%
\partial_{\chi}^{2}V|_{\text{vac}}=m_{\chi}^{2}=2\lambda_{\chi}\eta^{2}$, $%
\partial_{\varphi}^{2}V|_{\text{vac}}=m_{\varphi}^{2}=2\lambda_{\varphi
}\sigma^{2}$ with off-diagonal terms $\partial_{\chi\varphi}^{2}V|_{\text{vac%
}}=m_{\chi\varphi}^{2}=m_{\varphi\chi}^{2}=\pm2\beta\eta \sigma\equiv\mu^{2}$%
, which are nonvanishing for the case of interacting fields for which $%
\beta\neq0$. The meson mass (squared) matrix in terms of the flavor states $%
\chi$ and $\varphi$ is therefore given by 
\end{subequations}
\begin{equation}
\boldsymbol{M}^{2}=\left( 
\begin{array}{cc}
m_{\chi}^{2} & \mu^{2} \\ 
\mu^{2} & m_{\varphi}^{2}%
\end{array}
\right) =\left( 
\begin{array}{cc}
2\lambda_{\chi}\eta^{2} & \pm2\beta\eta\sigma \\ 
\pm2\beta\eta\sigma & 2\lambda_{\varphi}\sigma^{2}%
\end{array}
\right)  \label{5}
\end{equation}

(The sign of the off-diagonal terms in $\boldsymbol{M}^{2}$ are determined
by the signs of the vacuum states and the sign of $\beta$, i.e., $%
\mu^{2}=(\pm|\beta|)(\pm\eta)(\pm\sigma)=\pm|\beta|\eta\sigma$ with $\eta>0$
and $\sigma>0$.) The eigenmasses are given by%
\begin{equation}
m_{\pm}^{2}=\frac{1}{2}\left[ m_{\chi}^{2}+m_{\varphi}^{2}\pm\sqrt{4\mu
^{4}+(m_{\chi}^{2}-m_{\varphi}^{2})^{2}}\right]  \label{5a}
\end{equation}

indicating that the perturbative meson flavor states $\chi(x,t)$ and $%
\varphi(x,t)$ are not mass eigenstates, but rather, are linear combinations
of mass eigenstates $\phi_{+}(x,t)$ and $\phi_{-}(x,t)$:%
\begin{equation}
\left( 
\begin{array}{c}
\chi \\ 
\varphi%
\end{array}
\right) =\left( 
\begin{array}{cc}
\cos\theta & \sin\theta \\ 
-\sin\theta & \cos\theta%
\end{array}
\right) \left( 
\begin{array}{c}
\phi_{+} \\ 
\phi_{-}%
\end{array}
\right)  \label{5b}
\end{equation}

with $\theta$ a fixed \textquotedblleft mixing parameter\textquotedblright.
This mixing, to be discussed later, leads to oscillations of the flavor
meson states $\chi$ and $\varphi$ for $\beta\neq0$.

\section{Perturbative corrections}

\ \ A parameter $g$ is introduced to allow us to formally write the
potential in the form%
\begin{equation}
V(\chi,\varphi)=V_{0}(\chi,\varphi)+gV_{1}(\chi,\varphi)  \label{6}
\end{equation}

with $g$ being an expansion, or control, parameter such that $0\leq g\leq1$.
For $g=0$ we have the unperturbed potential $V_{0}$ and when $g=1$ we have
the full potential $V_{0}+V_{1}$. A set of correction equations for the
scalar fields can be obtained which are independent of $g$, and in
calculations involving $V_{1}$ we adopt the setting $g=1$. For now, however,
the value of $g$ is left arbitrary, but restricted to $g\in\lbrack0,1]$.

\bigskip

\ \ The functions $F(\chi,\varphi)$ and $G(\chi,\varphi)$ are defined as
derivatives of the potential $V(\chi,\varphi)$ with respect to the fields $%
\chi$ and $\varphi$, respectively:%
\begin{equation}
\begin{array}{cc}
F(\chi,\varphi)=F_{0}(\chi,\varphi)+F_{1}(\chi,\varphi)=\dfrac{\partial
V(\chi,\varphi)}{\partial\chi}=\partial_{\chi}V(\chi,\varphi) &  \\ 
G(\chi,\varphi)=G_{0}(\chi,\varphi)+G_{1}(\chi,\varphi)=\dfrac{\partial
V(\chi,\varphi)}{\partial\varphi}=\partial_{\varphi}V(\chi,\varphi) & 
\end{array}
\label{a6}
\end{equation}

where $F_{0}=\partial_{\chi}V_{0}$, $F_{1}=\partial_{\chi}V_{1}$, $%
G_{0}=\partial_{\varphi}V_{0}$, and $G_{1}=\partial_{\varphi}V_{1}$. The
quantities $F(\chi_{0},\varphi_{0})$ and $G(\chi_{0},\varphi_{0})$ etc. are
defined as $F$ and $G$ evaluated at $(\chi_{0},\varphi_{0})$,%
\begin{equation}
F(\chi_{0},\varphi_{0})=F(\chi,\varphi)\Big|_{\chi_{0},\varphi_{0}},\ \ \ \
\ G(\chi_{0},\varphi_{0})=G(\chi,\varphi)\Big|_{\chi_{0},\varphi _{0}}
\label{a6a}
\end{equation}

It is useful to introduce an abbreviated notation where the field $\psi$
denotes either $\chi$ or $\varphi$ and the function $H$ denotes either $F$
or $G$:%
\begin{equation}
\psi=\chi,\varphi,\ \ \ \ H(\psi)=F(\chi,\varphi),G(\chi,\varphi ),\ \ \ \ \
H(\psi)=H_{0}(\psi)+H_{1}(\psi)  \label{7}
\end{equation}

The equations of motion following from $\mathcal{L}$ are given by $\square
\psi+H(\psi)=0$, i.e.,%
\begin{equation}
\square\chi+F(\chi,\varphi)=0,\ \ \ \ \ \square\varphi+G(\chi,\varphi)=0
\label{8}
\end{equation}

given explicitly by (\ref{3}) for the potential (\ref{2}).

\bigskip

\ \ At this point a Rayleigh-Schr\"{o}dinger approach for obtaining \textit{%
static} corrections $\delta \psi (x)$ is implemented by writing $V(\chi
,\varphi )=V_{0}(\chi ,\varphi )+gV_{1}(\chi ,\varphi )$, so that $H(\psi
)=H_{0}(\psi )+gH_{1}(\psi )$. (However, it must be stated that, more
generally, one expects \textit{nonstatic} corrections to exist, since, as
will be seen later, the interaction $V_{1}$ will result in interkink forces
between the $\chi $ and $\varphi $ kinks, allowing a relative motion between
them. Nevertheless, we adopt a \textquotedblleft
quasistatic\textquotedblright\ type of approach where, for the purpose of
simplifications, the time dependence of $\delta \psi $ is neglected. We
justify this on the basis that the interaction control parameter $\beta $ is
small, i.e., $|\beta |\ll \lambda _{\chi }$,$\lambda _{\varphi }$, resulting
in relatively weak interkink forces.) When $gH_{1}=0$ we have the
unperturbed system, described by the unperturbed solution $\psi _{0}(x)$,
obeying $\square \psi _{0}+H_{0}(\psi _{0})=0$. However, for $gH_{1}\neq 0$
the full solution $\psi =\psi _{0}+\delta \psi $ has a dependence upon the
parameter $g$, i.e., $\psi =\psi (x,g)=\psi _{0}(x)+\delta \psi (x,g)$. We
assume that $H_{1}$ is a small perturbation and that $\delta \psi $ is
dominated by the base solution $\psi _{0}$ ( specifically, $|\delta \psi
|\ll |\psi _{\text{vac}}|$, where $|\psi _{\text{vac}}|=\eta $ or $\sigma $%
). \ The correction $\delta \psi $ due to the perturbation $gH_{1}$ can then
be expanded in powers of $g$ as in the case of the Rayleigh-Schr\"{o}dinger
method in quantum mechanics,%
\begin{equation}
\begin{array}{ll}
\psi (x,g) & =\psi _{0}(x)+\delta \psi (x,g) \\ 
\delta \psi (x,g) & =\sum_{n=1}^{\infty }g^{n}\psi _{n}(x)=g\psi
_{1}(x)+g^{2}\psi _{2}(x)+\cdot \cdot \cdot%
\end{array}
\label{9}
\end{equation}

with $\psi=\chi,\varphi$.

\bigskip

\ \ Next, we can expand the potential $V(\psi)$ and its derivatives $%
H(\psi)=\partial_{\psi}V(\psi)$ about the base solution $\psi_{0}$. We then
have%
\begin{equation}
H(\chi,\varphi)=H(\psi_{0})+\left( \delta\chi\partial_{\chi}+\delta
\varphi\partial_{\varphi}\right) H(\chi,\varphi)\Big|_{\chi_{0},\varphi_{0}}+%
\frac{1}{2}\left( \delta\chi\partial_{\chi}+\delta\varphi\partial_{\varphi
}\right) ^{2}H(\chi,\varphi)\Big|_{\chi_{0},\varphi_{0}}+\cdot\cdot \cdot
\label{10}
\end{equation}

where $(\delta\psi\partial_{\psi})^{2}H=(\delta\psi)^{2}\partial_{\psi}^{2}H$
and so on. Since $H=H_{0}+gH_{1}$ this becomes%
\begin{align}
H(\chi,\varphi) & =H_{0}(\psi_{0})+\left(
\delta\chi\partial_{\chi}+\delta\varphi\partial_{\varphi}\right)
H_{0}(\chi,\varphi)\Big|_{\chi _{0},\varphi_{0}}+\frac{1}{2}\left(
\delta\chi\partial_{\chi}+\delta \varphi\partial_{\varphi}\right)
^{2}H_{0}(\chi,\varphi)\Big|_{\chi _{0},\varphi_{0}}+\cdot\cdot\cdot  \notag
\\
& +gH_{1}(\psi_{0})+g\left( \delta\chi\partial_{\chi}+\delta\varphi
\partial_{\varphi}\right) H_{1}(\chi,\varphi)\Big|_{\chi_{0},\varphi_{0}}+%
\frac{1}{2}g\left( \delta\chi\partial_{\chi}+\delta\varphi\partial
_{\varphi}\right) ^{2}H_{1}(\chi,\varphi)\Big|_{\chi_{0},\varphi_{0}}+\cdot%
\cdot\cdot  \label{11}
\end{align}

\bigskip

\ \ The equations of motion for the full system (\ref{3}) can now be written
in expanded form \ with the aid of (\ref{9}) and (\ref{11}) (See Appendix
A.). We will confine attention to first order corrections, for which $\psi
(x)=\psi_{0}(x)+\psi_{1}(x)$, where the corrections for $\psi_{1}$ are given
by 
\begin{subequations}
\label{12}
\begin{align}
\square\chi_{1}+(\chi_{1}\partial_{\chi}+\varphi_{1}\partial_{%
\varphi})F_{0}(\chi_{0},\varphi_{0})+F_{1}(\chi_{0},\varphi_{0}) & =0
\label{12a} \\
\square\varphi_{1}+(\chi_{1}\partial_{\chi}+\varphi_{1}\partial_{\varphi
})G_{0}(\chi_{0},\varphi_{0})+G_{1}(\chi_{0},\varphi_{0}) & =0  \label{12b}
\end{align}

For our model given by (\ref{1}) and (\ref{2}), 
\end{subequations}
\begin{equation}
\begin{array}{lll}
F_{0}(\chi,\varphi)=\lambda_{\chi}\chi(\chi^{2}-\eta^{2}), &  & 
F_{1}(\chi,\varphi)=\beta\chi(\varphi^{2}-\sigma^{2}) \\ 
G_{0}(\chi,\varphi)=\lambda_{\varphi}\varphi(\varphi^{2}-\sigma^{2}), &  & 
G_{1}(\chi,\varphi)=\beta\varphi(\chi^{2}-\eta^{2})%
\end{array}
\label{15}
\end{equation}

\section{Non-interacting kinks, $\boldsymbol{\protect\beta=0}$}

\ \ We consider 1-dimensional (1D) domain kinks and/or \textquotedblleft
planar\textquotedblright, or \textquotedblleft flat\textquotedblright, $N$%
-dimensional ($N$D) domain walls localized in the $x$ direction. Although
the domain defects can generally be dynamic, with 1D kinks moving along the $%
x$ axis, and the $N$D defects being able to translate and wiggle. Other
types of dynamical motions are also possible. However, the focus here will
be primarily upon static configurations that depend upon the single
coordinate $x$.

\bigskip

\ \ For the case where there is no interaction between the $\chi$ and $%
\varphi$ fields, $\beta=0$ and therefore $V_{1}=0$. In this case the
potential is simply $V=V_{0}$, with vacuum states and masses given by $%
|\psi_{\text{vac}}|=v$, where $v=$ $\eta$ or $\sigma$, and the perturbative
meson masses given in (\ref{5}), which might be written symbolically as $%
m^{2}=2\lambda v^{2}$. The nonperturbative, topological (kink) solutions for 
$\beta=0$ satisfying (\ref{3}), i.e., $-\partial_{x}^{2}\psi_{0}+\lambda
\psi_{0}(\psi_{0}^{2}-v^{2})=0$, are given in abbreviated form by 
\begin{equation}
\begin{array}{ll}
\psi_{0}(x)=v\tanh\left[ \kappa(x-x_{0})\right] =v\tanh\left[ \dfrac{%
(x-x_{0})}{w}\right] &  \\ 
\kappa=\dfrac{1}{w}=\sqrt{\dfrac{\lambda}{2}}v,\ \ \ w=2\delta=\sqrt{\dfrac {%
2}{\lambda}}\dfrac{1}{v}=\dfrac{2}{m},\ \ \ m=\sqrt{2\lambda}v & 
\end{array}
\label{16}
\end{equation}

where $v$ represents either $\eta$ or $\sigma$, $x_{0}$ is the position of
the kink/wall, $w$ is its width parameter (i.e., length along the $x$ axis),
and $\delta=m^{-1}$ is the half-width parameter. The parameter $\kappa =1/w=%
\frac{1}{2}m$ is the inverse of the width parameter, or half of the
(perturbative) particle mass $m$. For the $\chi$ and $\varphi$ kinks we
write specifically,

\begin{subequations}
\label{e17}
\begin{align}
& 
\begin{array}{ll}
\chi_{0}(x)=\eta\tanh\left[ k_{\chi}(x-x_{\chi})\right] =\eta\tanh\left( 
\dfrac{x-x_{\chi}}{w_{\chi}}\right) , &  \\ 
k_{\chi}=\dfrac{1}{w_{\chi}}=\sqrt{\dfrac{\lambda_{\chi}}{2}}\eta=\frac{1}{2}%
m_{\chi},\ \ \ w_{\chi}=\dfrac{1}{\eta}\sqrt{\dfrac{2}{\lambda_{\chi}}}, & 
\end{array}
\label{e17a} \\
&  \notag \\
& 
\begin{array}{ll}
\varphi_{0}(x)=\sigma\tanh\left[ k_{\varphi}(x-x_{\varphi})\right]
=\sigma\tanh\left( \dfrac{x-x_{\varphi}}{w_{\varphi}}\right) , &  \\ 
k_{\varphi}=\dfrac{1}{w_{\varphi}}=\sqrt{\dfrac{\lambda_{\varphi}}{2}}\sigma=%
\frac{1}{2}m_{\varphi},\ \ \ w_{\varphi}=\dfrac{1}{\sigma}\sqrt {\dfrac{2}{%
\lambda_{\varphi}}}, & 
\end{array}
\label{e17b}
\end{align}

where $x_{\chi}$ and $x_{\varphi}$ are the positions of the $\chi$ and $%
\varphi$ kinks with widths (i.e., lengths along the $x$ axis) of $w_{\chi}$
and $w_{\varphi}$. Antikink ($\bar{\psi}_{0}$) solutions are given by $\bar{%
\chi}_{0}=-\chi_{0}$ and $\bar{\varphi}_{0}=-\varphi_{0}$. Time-dependent
Lorentz boosted kink solutions are given by 
\end{subequations}
\begin{equation}
\begin{array}{cc}
\chi_{0}(x,t) & =\eta\tanh\left[ \dfrac{\left( x-x_{\chi}\right) -u_{\chi }t%
}{w_{\chi}(1-u_{\chi}^{2})^{1/2}}\right] \smallskip \\ 
\varphi_{0}(x,t) & =\sigma\tanh\left[ \dfrac{\left( x-x_{\varphi}\right)
-u_{\varphi}t}{w_{\varphi}\left( 1-u_{\varphi}^{2}\right) ^{1/2}}\right]%
\end{array}
\label{20}
\end{equation}

where $u_{\chi}$, $u_{\varphi}$ are the kink velocities. These represent two
ordinary, non-interacting kinks, which can freely pass through one another
on the $x$ axis. For static kinks, $\chi$ and $\varphi$ rapidly enter their
respective vacuum states, i.e., $\chi\rightarrow\pm\eta$ and $\varphi
\rightarrow\pm\sigma$ for $|x-x_{\chi}|\gg w_{\chi}$ and $|x-x_{\varphi}|\gg
w_{\varphi}$, with each kink or antikink interpolating between the two
vacua. (Note that the $\chi$ and $\varphi$ kink solutions approach vacuum
states quite rapidly for $|x-x_{0}|\gtrsim2w$.) In general, multiple kinks
and antikinks can exist along the $x$ axis, and $\chi$ and $\varphi$ K-\={K}
annihilations can produce $\chi$ and $\varphi$ bosons, respectively, in the
process.

\section{Interacting defects: $\mathbf{\protect\beta\neq0}$}

\ \ The form of the first order equations for the correction $\psi _{1}(x)$
from (\ref{12}) is given by%
\begin{equation}
-\partial _{x}^{2}\psi _{1}+(\chi _{1}\partial _{\chi }+\varphi _{1}\partial
_{\varphi })H_{0}(\psi _{0})+H_{1}(\psi _{0})=0  \label{21}
\end{equation}

With the help of (\ref{15}) and (\ref{e17}) we have for $H_{0}$ and $H_{1}$
terms%
\begin{equation}
\begin{array}{lll}
F_{0}(\chi_{0},\varphi_{0}) & =\lambda_{\chi}\chi_{0}(\chi_{0}^{2}-\eta^{2})
& =-\lambda_{\chi}\eta^{3}\tanh k_{\chi}(x-x_{\chi})\text{sech}^{2}k_{\chi
}(x-x_{\chi}) \\ 
G_{0}(\chi_{0},\varphi_{0}) & =\lambda_{\varphi}\varphi_{0}(\varphi_{0}^{2}-%
\sigma^{2}) & =-\lambda_{\varphi}\sigma^{3}\tanh k_{\varphi}(x-x_{\varphi})%
\text{sech}^{2}k_{\varphi}(x-x_{\varphi}) \\ 
F_{1}(\chi_{0},\varphi_{0}) & =\beta\chi_{0}(\varphi_{0}^{2}-\sigma^{2}) & 
=-\beta\eta\sigma^{2}\tanh k_{\chi}(x-x_{\chi})\text{sech}^{2}k_{\varphi
}(x-x_{\varphi}) \\ 
G_{1}(\chi_{0},\varphi_{0}) & =\beta\varphi_{0}(\chi_{0}^{2}-\eta^{2}) & 
=-\beta\eta^{2}\sigma\tanh k_{\varphi}(x-x_{\varphi})\text{sech}^{2}k_{\chi
}(x-x_{\chi})%
\end{array}
\label{22}
\end{equation}

where we make use of the identity $\tanh^{2}u-1=-$sech$^{2}u$. In addition,%
\begin{equation}
\begin{array}{lll}
\partial_{\chi}F_{0}(\chi_{0},\varphi_{0}) & =\lambda_{\chi}(3\chi_{0}^{2}-%
\eta^{2}) & =\lambda_{\chi}\eta^{2}\left[ 3\tanh^{2}k_{\chi}(x-x_{\chi })-1%
\right] \\ 
\partial_{\varphi}G_{0}(\chi_{0},\varphi_{0}) & =\lambda_{\varphi}(3%
\varphi_{0}^{2}-\sigma^{2}) & =\lambda_{\varphi}\sigma^{2}\left[ 3\tanh
^{2}k_{\varphi}(x-x_{\varphi})-1\right] \\ 
\partial_{\varphi}F_{0}(\chi_{0},\varphi_{0}) & =0 &  \\ 
\partial_{\chi}G_{0}(\chi_{0},\varphi_{0}) & =0 & 
\end{array}
\label{23}
\end{equation}

\ \ We now choose to set the $\chi$ kink to be located at the origin, $%
x_{\chi}=0$, and the $\varphi$ kink to be located at $x_{\varphi}=a\geq0$.
Then by (\ref{12}) and (\ref{21})-(\ref{23}) the equations for the first
order corrections for the static fields become 
\begin{subequations}
\label{24}
\begin{align}
\chi_{1}^{\prime\prime}(x)-2k_{\chi}^{2}\left[ 3\tanh^{2}k_{\chi}x-1\right]
\chi_{1}(x) & =-\beta\eta\sigma^{2}\tanh k_{\chi}x\cdot\text{sech}%
^{2}k_{\varphi}(x-a)  \label{24a} \\
\varphi_{1}^{\prime\prime}(x)-2k_{\varphi}^{2}\left[ 3\tanh^{2}k_{\varphi
}(x-a)-1\right] \varphi_{1}(x) & =-\beta\eta^{2}\sigma\tanh k_{\varphi
}(x-a)\cdot\text{sech}^{2}k_{\chi}x  \label{24b}
\end{align}

where $2k_{\chi}^{2}=\lambda_{\chi}\eta^{2}$, $2k_{\varphi}^{2}=\lambda
_{\varphi}\sigma^{2}$ and $^{\prime}$ denotes differentiation with respect
to $x$.

\bigskip

\ \ The zeroth order antikink fields $\bar{\chi}_{0}$ and $\bar{\varphi}_{0}$
are given by $\bar{\chi}_{0}(x)=-\chi_{0}(x)$ and $\bar{\varphi}%
_{0}(x)=-\varphi_{0}(x)$, so that for first order corrections 
\end{subequations}
\begin{equation*}
\bar{\chi}(x)=-\chi_{0}(x)+\bar{\chi}_{1}(x),\ \ \ \ \ \bar{\varphi }%
(x)=-\varphi_{0}(x)+\bar{\varphi}_{1}(x)
\end{equation*}

The equations for the first order corrections $\bar{\chi}_{1}(x)$ and $\bar{%
\varphi}_{1}(x)$ are then obtained from (\ref{24}) by making replacements $%
\psi_{0}(x)\rightarrow\bar{\psi}_{0}(x)=-\psi_{0}(x)$, or $%
k(x-x_{0})\rightarrow-k(x-x_{0})$, i.e., $\tanh k(x-x_{0})\rightarrow-\tanh
k(x-x_{0})$, resulting in 
\begin{subequations}
\label{25}
\begin{align}
\bar{\chi}_{1}^{\prime\prime}(x)-2k_{\chi}^{2}\left[ 3\tanh^{2}k_{\chi }x-1%
\right] \bar{\chi}_{1}(x) & =+\beta\eta\sigma^{2}\tanh k_{\chi}x\cdot\text{%
sech}^{2}k_{\varphi}(x-a)  \label{25a} \\
\bar{\varphi}_{1}^{\prime\prime}(x)-2k_{\varphi}^{2}\left[ 3\tanh
^{2}k_{\varphi}(x-a)-1\right] \bar{\varphi}_{1}(x) &
=+\beta\eta^{2}\sigma\tanh k_{\varphi}(x-a)\cdot\text{sech}^{2}k_{\chi}x
\label{25b}
\end{align}

A comparison of (\ref{24}) and (\ref{25}) implies that $\bar{\psi}%
_{1}(x)=-\psi_{1}(x)$. Note that for the equations for the first order
corrections the right hand sides depend upon the functions $F_{1}(\chi
_{0},\varphi_{0})$ and $G_{1}(\chi_{0},\varphi_{0})$, and therefore upon the
form of the interaction chosen for $V_{1}(\chi,\varphi)$.

\section{Approximate first order corrections}

\subsection{\textquotedblleft Thin wall\textquotedblright\ (delta function)
approximation}

\ \ Exact analytic solutions of the DEs of (\ref{24}) have proven to be
rather evasive, as they involve different hyperbolic functions with
different arguments. Instead, \textit{approximate} analytic representations
of the solutions have been found, by using a type of \textquotedblleft%
\textit{thin wall}\textquotedblright\ approximation for the kinks/walls
where a sech$^{2}$ function is approximated by a Dirac delta function, each
of which has a \textquotedblleft sifting\textquotedblright\ property. This
approximation allows the DEs to be rewritten and solved with much greater
ease with the techniques commonly used in quantum mechanical problems with
delta function potentials.

\bigskip

\ \ There exist many representations of a Dirac delta function $\delta(x)$
in terms of limiting forms of well defined functions. One such
representation can be written in terms of the sech$^{2}$ function.
Specifically (see, e.g., \cite{Wolfram}) , 
\end{subequations}
\begin{equation}
\delta(x)=\lim_{k\rightarrow\infty}\left( \frac{1}{2}k\text{ sech}%
^{2}kx\right) =\lim_{w\rightarrow0}\left( \frac{1}{2}\frac{1}{w}\text{ sech}%
^{2}\frac{x}{w}\right) =\left\{ 
\begin{array}{cc}
\infty, & x=0 \\ 
0, & x\neq0%
\end{array}
\right\} ,\ \text{with}\ \int_{-\infty}^{\infty}\delta(x)dx=1  \label{26}
\end{equation}

For a high and narrow function $k\cdot$sech$^{2}kx$ (with \textquotedblleft
width\textquotedblright\ parameter $w=1/k$), we expect the function $\frac {1%
}{2}k$ sech$^{2}kx$ to exhibit similar \textquotedblleft
sifting\textquotedblright\ properties as a delta function. The
nonhomogeneous DEs of (\ref{24}) can be modified and solved \textit{%
approximately} if we use the approximation%
\begin{equation}
\text{sech}^{2}kx\rightarrow\frac{2}{k}\delta(x)  \label{27}
\end{equation}

This approximation allows the sech$^{2}$ function to have the simple sifting
property of a delta function, while holding the parameter $k$ finite. The
approximation is expected to become better for larger $k$, but even for
smaller values of $k$ we should see fundamental features of a solution in an
analytic form where the roles of the various parameters of the system are
shown explicitly. These parameters can be important in subsequent
calculations.

\subsection{Approximate Solutions}

\ \ \textbf{The }$\boldsymbol{\chi}_{1}$ \textbf{correction:}\ \ \ For
brevity we temporarily denote $\chi_{1}$ by $\chi_{1}(x)=\psi(x)$, and adopt
the settings $k_{\chi}=k$, $k_{\varphi}=q$, $x_{\chi}=0$, and $x_{\varphi}=a$%
. The location of the $\chi$ kink is $x=0$, and that of the $\varphi$ kink
is $x=a$. Also define the constant $B_{1}=\beta\eta\sigma^{2}$. Then (\ref%
{24a}) is given by%
\begin{equation}
\psi^{\prime\prime}(x)-2k^{2}\left[ 3\tanh^{2}kx-1\right] \psi
(x)=-B_{1}\tanh kx\text{ sech}^{2}q(x-a)  \label{28}
\end{equation}

where the prime denotes differentiation with respect to $x$. Using the
identity $\tanh^{2}-1=-$sech$^{2}$, we have $[3\tanh^{2}-1]=[-3$ sech$%
^{2}+2] $. Therefore (\ref{28}) can be rewritten as%
\begin{equation}
\psi^{\prime\prime}-4k^{2}\psi+(6k^{2}\text{sech}^{2}kx)\psi=-B_{1}\tanh kx%
\text{ sech}^{2}q(x-a)  \label{29}
\end{equation}

We now assume that the kinks are sufficiently narrow to make the delta
function approximations%
\begin{equation}
\text{sech}^{2}kx\rightarrow\frac{2}{k}\delta(x),\ \ \ \ \ \text{sech}%
^{2}q(x-a)\rightarrow\frac{2}{q}\delta(x-a)  \label{30}
\end{equation}

although $k$ and $q$ are kept large, but finite, so that each of the sech$%
^{2}$ functions has a very narrow, but finite width, and has a large, but
finite height. The sech$^{2}$ functions are finite, but sufficiently highly
peaked and narrow that we use the delta functions as rough approximations.

\bigskip

\ \ We therefore have the approximate second order nonhomogeneous
differential equation (DE)%
\begin{equation}
\psi^{\prime\prime}(x)-4k^{2}\psi(x)+12k\delta(x)\psi(x)=-\frac{2B_{1}}{q}%
\tanh kx\cdot\delta(x-a)  \label{31}
\end{equation}

The delta function approximation has introduced discontinuities at $x=0$ and 
$x=a$. We require that $\psi(x)$ be continuous, and following the procedure
used in quantum mechanics we integrate the DE in small neighborhoods about $%
x=0$ and $x=a$ to obtain $\psi^{\prime}(0)$ and $\psi^{\prime}(a)$. Due to
the two discontinuities, we divide the $x$ space into three continuous
regions: region I, $x<0$, region II, $0<x<a$, and region III, $x>a$. In each
delta function-free region, we have the same DE, namely, $\psi^{\prime\prime
}-4k^{2}\psi=0$ with exponential solutions $e^{\pm2kx}$. The boundary
conditions are $\psi\rightarrow0$ as $x\rightarrow\pm\infty$. We then have
the solutions%
\begin{equation}
\begin{array}{ll}
\text{I:\ \ \ }\psi_{1}=Ae^{2kx}, & x<0 \\ 
\text{II: \ }\psi_{2}=Be^{2kx}+Ce^{-2kx}, & 0<x<a \\ 
\text{III: }\psi_{3}=De^{-2kx}, & x>a%
\end{array}
\label{32}
\end{equation}

\ \ The continuity of $\psi$ at $x=0$ and $x=a$ and expressions for $%
\psi^{\prime}(0)$ and $\psi^{\prime}(a)$ allow the determination of the
constants $A,B,C,$ and $D$ (see Appendix B). The resulting solution is given
by (see FIG. 1) 
\begin{equation}
\chi_{1}(x)\approx\frac{\beta\sigma}{\sqrt{\lambda_{\chi}\lambda_{\varphi}}}%
e^{-2ka}\tanh ka\times\left\{ 
\begin{array}{ll}
-\frac{1}{2}e^{2kx}, & x<0 \\ 
e^{2kx}-\frac{3}{2}e^{-2kx}, & 0<x<a \\ 
\left( e^{4ka}-\frac{3}{2}\right) e^{-2kx}, & x>a%
\end{array}
\right\}  \label{38}
\end{equation}

\begin{figure}[tbh]
\centering
\includegraphics[width=8.0cm]{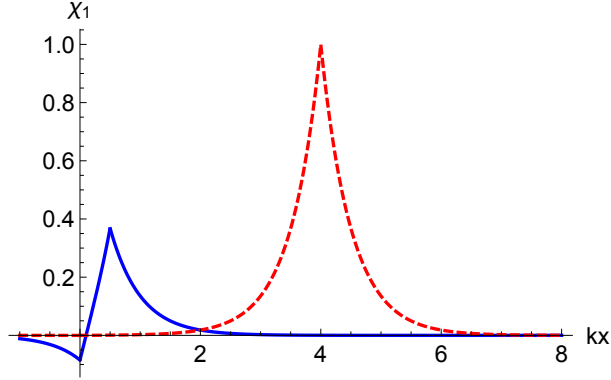}
\caption{$\protect\chi_{1}(x)$ vs $kx$ with $ka=.5$ (solid) and $ka=4$
(dashed). $B_{1}/2kq$ has been set to 1. Note how the $\protect\chi_{1}$
condensate appears at the position of the $\protect\varphi$ kink ($x=a$).}
\end{figure}

We have not included the \textquotedblleft zero mode\textquotedblright\
solution \cite{Goldstone75},\cite{Jackiw77} $\chi_{1}^{(0)}(x)\propto
\chi_{0}^{\prime}(x)\sim$ sech$^{2}kx$ of the homogeneous (i.e., sourceless)
DE, as this zero mode does not arise in response to the $\chi-\varphi$
interaction, and the solution of interest here, and points forward, is that
of (\ref{38}), which does arise from the two-kink interaction.

\bigskip

\ \ \textbf{The }$\boldsymbol{\varphi}_{1}$ \textbf{correction:}\ \ Now
denote $\varphi_{1}$ by $\varphi_{1}(x)=\psi(x)$, and again $k_{\chi}=k$, $%
k_{\varphi}=q$, $x_{\chi}=0$, and $x_{\varphi}=a$. The location of the $\chi$
kink is $x=0$, and that of the $\varphi$ kink is $x=a$, as before. Also
define the constant $B_{2}=\beta\eta^{2}\sigma$. Using the same
approximations as before, we divide the $x$ space into three regions with
functions $\psi _{1}(x)$, $\psi_{2}(x)$, and $\psi_{3}(x)$ in regions I, II,
and III, respectively. With the delta function approximation, (\ref{24b}) is
written as%
\begin{equation}
\psi^{\prime\prime}(x)-4q^{2}\psi(x)+12q\delta(x-a)\psi(x)=-\frac{2B_{2}}{k}%
\tanh q(x-a)\delta(x)  \label{41}
\end{equation}

with boundary conditions $\psi\rightarrow0$ as $x\rightarrow\pm\infty$. Each
region is again $\delta$ function-free, and the solutions are again of
exponential form $e^{\pm2qx}$. Specifically,%
\begin{equation}
\begin{array}{ll}
\text{I:\ \ \ }\psi_{1}=Ae^{2qx}, & x<0 \\ 
\text{II: \ }\psi_{2}=Be^{2qx}+Ce^{-2qx}, & 0<x<a \\ 
\text{III: }\psi_{3}=De^{-2qx}, & x>a%
\end{array}
\label{42}
\end{equation}

where the coefficients $A,B,C,D$ are now new ones for the $\varphi_{1}$
function. We use continuity of $\psi(x)$ at $x=0$ and $x=a,$ and integrate
the DE (\ref{41}) to obtain constraints on $\psi^{\prime}(0)$ and $%
\psi^{\prime }(a)$. The coefficients can be determined (see Appendix B) and
the resulting solution is given by (see FIG. 2)%
\begin{equation}
\varphi_{1}(x)\approx-\frac{\beta\eta}{\sqrt{\lambda_{\chi}\lambda_{\varphi}}%
}\tanh qa\times\left\{ 
\begin{array}{ll}
\left( 1-\frac{3}{2}e^{-4qa}\right) e^{2qx}, & x<0 \\ 
-\frac{3}{2}e^{-4qa}e^{2qx}+e^{-2qx}, & 0<x<a \\ 
-\frac{1}{2}e^{-2qx}, & x>a%
\end{array}
\right\}  \label{47}
\end{equation}

\begin{figure}[tbh]
\centering
\includegraphics[width=8.0cm]{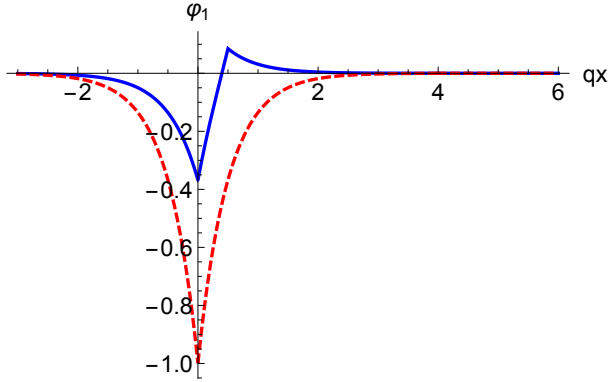}
\caption{$\protect\varphi_{1}(x)$ vs $qx$ for $qa=.5$ (solid) and $qa=4$
(dashed). $B_{2}/2kq$ has been set to 1. Note how the $\protect\varphi_{1}$
condensate appears at the position of the $\protect\chi$ kink ($x=0$).}
\label{fig2}
\end{figure}

\ \ Once again, there is a zero mode solution \cite{Goldstone75},\cite%
{Jackiw77}, $\varphi_{1}^{(0)}(x)\propto\varphi_{0}^{\prime}(x)\sim$ sech$%
^{2}q(x-a)$ which solves the homogeneous (sourceless) DE of\ (\ref{24b}),
but since it has nothing to do with the $\chi-\varphi$ interaction we
dismiss it from further consideration.

\ \ 

\ \ Note that for $a\neq0$ (\ref{38}) and (\ref{47}) suggest that the
correction for \textit{each} kink/wall solution manifests itself in the form
of a \textquotedblleft ghostly\textquotedblright\ displaced scalar
condensate, residing within the \textit{other} kink/wall. For example, the $%
\chi_{1}$ correction is pronounced near $x=a$ (the location of the $\varphi$
kink), and the $\varphi_{1}$ correction is pronounced near $x=0$ (the
location of the $\chi$ kink). (From (\ref{24}) it is seen that a correction $%
\psi_{1}(x)$ for either kink can not vanish at the location of the other%
\textbf{\ }kink, as the source term for each correction maximizes at the
location of the other kink.) Therefore, the $\chi$ kink has a topological
structure from the $\chi$ field, along with a condensate from the $\varphi$
field, and vice versa. The kink, along with the condensate within it, might
be referred to as a \textquotedblleft structured\textquotedblright\ kink.
The condensates described by $\chi_{1}$ and $\varphi_{1}$ vanish for $a=0$,
i.e., when the centers of the kinks coincide. Therefore, within either
kink/wall there appears a small additional energy density due to the
condensate when there is a nonzero separation between them ($a\neq0$).
However, this extra mass disappears when the two kinks overlap, suggesting
the existence of a weakly bound state under certain circumstances.

\section{Structured solitons}

\subsection{Displaced condensates}

\ \ The $\chi_{0}(x)$ and $\varphi_{0}(x)$ static kinks are located at $x=0$
and $x=a$, respectively. The excitation modes $\chi_{1}(x)$ and $\varphi
_{1}(x)$ are described by (\ref{38}) and (\ref{47}), and \textit{each}
exhibits an enhancement, or scalar field condensate, at the position of the%
\textit{\ other} kink. Specifically, the $\chi_{1}$ mode is concentrated at $%
x=a$, the location of the $\varphi_{0}$ kink, and the $\varphi_{1}$ mode is
concentrated at at $x=0$, the location of the $\chi_{0}$ kink. The
\textquotedblleft widths\textquotedblright\ of the condensates are
comparable to, or on the order of, those of the host kinks.

\bigskip

\ These condensate modes are nontopological in nature, as each mode solution
rapidly approaches its asymptotic value of zero. However, a condensate has
an attendant \textquotedblleft mass\textquotedblright\ (surface energy, for
a domain wall) that we can denote by $\Sigma$. This mass is obtained from
the energy-momentum tensor $T_{\mu\nu}^{\psi}$ associated with the
condensate; specifically, $\Sigma_{\psi}=\int T_{00}^{\psi}(x)dx$ for each
condensate mode ($\psi=\chi$ or $\varphi$). We consider the kinks to be
separated by a distance $a$, and assume, for simplicity, that $%
\chi_{0}\approx\eta$ at the location of $\varphi_{0}$ (i.e., at $x\sim a>0$)
and $\varphi_{0}\approx-\sigma$ at the location of $\chi_{0}$ (i.e., at $%
x\sim0$). The kink separation is considered to be on the order of, or
greater than, the kink \textquotedblleft widths\textquotedblright, $a\gtrsim
w_{\chi},w_{\varphi}$, although with some justification we will be able to
extrapolate our results for the case where $a\rightarrow0$.

\subsection{\textquotedblleft Masses\textquotedblright\ of the condensates}

\ \ The basic idea used here is to isolate the contribution of each
condensate ($\psi_{1}(x)$) to the energy-momentum $T_{\mu\nu}^{\psi}(x)$ and
then integrate $T_{00}^{\psi}(x)$ to obtain the mass $\Sigma_{\psi}(a)=\int
T_{00}^{\psi}(x)dx$ which will depend upon the separation distance $a$
between the $\chi_{0}$ and $\varphi_{0}$ kinks. The computational details
are given in Appendix C, and we simply state the results here for $%
\Sigma_{\chi}$ and $\Sigma_{\varphi}$, i.e., the masses of the $\chi_{1}$
and $\varphi_{1}$ condensates, respectively:\ \ 
\begin{subequations}
\label{e48}
\begin{align}
\Sigma_{\chi}(a) & \approx2\beta^{2}\eta\sigma^{2}\left[ \sqrt {\frac{%
\lambda_{\chi}}{2}}\frac{\tanh^{2}ka}{\lambda_{\chi}\lambda_{\varphi}}+\sqrt{%
\frac{2}{\lambda_{\varphi}}}\frac{\tanh ka}{\sqrt{\lambda_{\chi
}\lambda_{\varphi}}}\right]  \label{e48a} \\
\Sigma_{\varphi}(a) & \approx2\beta^{2}\eta^{2}\sigma\left[ \sqrt {\frac{%
\lambda_{\varphi}}{2}}\frac{\tanh^{2}qa}{\lambda_{\chi}\lambda _{\varphi}}+%
\sqrt{\frac{2}{\lambda_{\chi}}}\frac{\tanh qa}{\sqrt{\lambda
_{\chi}\lambda_{\varphi}}}\right]  \label{e48b}
\end{align}

\ \ The results for the \textquotedblleft masses\textquotedblright\ $%
\Sigma_{\chi}$ and $\Sigma_{\varphi}$, given by (\ref{e48}) allow us to
reasonably expect that each mass decreases with decreasing separation
distance $a$, presumably to zero when the centers of the two kinks coincide.
(This expectation is strengthened by noticing that the corrections $\chi_{1}$
and $\varphi_{1}$ vanish as $a\rightarrow0$.) Such a decrease in the total
system mass suggests the presence of a weak ($\propto\beta^{2}$), but
nonzero, force of attraction between the two kinks, allowing a weakly bound
state to exist. This attractive force must be of fairly short range, since $%
\tanh\kappa a$ approaches unity for $\kappa a\sim2$, where $\kappa=1/w$ is
the inverse width parameter ($\kappa=k=1/w_{\chi}$ or $\kappa=q=1/w_{%
\varphi} $).

\subsection{Weakly bound states}

\ \ A structured $\chi$ soliton resides at $x=0$, comprised of the $\chi_{0}$
topological kink and the $\varphi_{1}$ condensate. Likewise, a structured $%
\varphi$ soliton resides at $x=a$, comprised of the $\varphi_{0}$
topological kink and the $\chi_{1}$ condensate. Each topological kink has an
energy density of the form $T_{00}^{\psi_{0}}(x)=\frac{1}{2}\lambda v^{4}$%
sech$^{4}\kappa x$, where $v=\eta$ or $\sigma$, $\lambda=\lambda_{\chi}$ or $%
\lambda_{\varphi}$, and $\kappa=k$ or $q$. The \textquotedblleft
masses\textquotedblright\ of the topological kinks are 
\end{subequations}
\begin{equation}
M_{\chi}=\frac{2}{3}\sqrt{2\lambda_{\chi}}\eta^{3},\ \ \ \ \ M_{\varphi}=%
\frac{2}{3}\sqrt{2\lambda_{\varphi}}\sigma^{3}  \label{67}
\end{equation}

Therefore, the total \textquotedblleft masses\textquotedblright\ of the
structured solitons with condensates are%
\begin{equation}
\mu_{\chi}(a)=M_{\chi}+\Sigma_{\varphi}(a),\ \ \ \ \ \mu_{\varphi
}(a)=M_{\varphi}+\Sigma_{\chi}(a)  \label{68}
\end{equation}

with $\Sigma_{\chi}$ and $\Sigma_{\varphi}$ given by (\ref{e48}). (The
\textquotedblleft masses\textquotedblright\ $\mu_{\chi}$, $\mu_{\varphi}$
have dimensions of mass for $1+1$ dimensional kinks or mass$^{3}$ for $3+1$
dimensional domain walls.)

\bigskip

\ \ It has been suggested that when the two structured solitons are at the
same position, $a\rightarrow0$, then the condensate masses vanish, $%
\Sigma_{\chi}\rightarrow0$ and $\Sigma_{\varphi}\rightarrow0$. This
suggestion is strengthened by noticing from (\ref{38}) and (\ref{47}) that $%
\chi_{1}\rightarrow0$ and $\varphi_{1}\rightarrow0$ as $a\rightarrow0$,
without an assumption that $\kappa a\gtrsim1$. We therefore expect the
corresponding energy densities $T_{00}^{\chi,\varphi}$ to vanish as $%
a\rightarrow0$, i.e., $\Sigma_{\chi}\rightarrow0$, $\Sigma_{\varphi
}\rightarrow0$ as $a\rightarrow0$. \ So when the two structured solitons
coincide at the same position, the mass of each decreases so that there is a
\textquotedblleft mass defect\textquotedblright\ of the two-soliton system%
\begin{equation}
\Delta\mu=\mu_{\text{Total,max}}-M_{\text{Total}}=\left(
\Sigma_{\chi}+\Sigma_{\varphi}\right) _{\text{max}}  \label{69}
\end{equation}

where $\Sigma_{\text{max}}$ is the maximum value of $\Sigma$, evaluated for $%
\tanh\kappa a=1$ ($\kappa=k$ or $q)$. This mass defect, or binding energy,
is the amount of energy required to separate a two-soliton bound state at
rest into two separate solitons. We surmise that the structured solitons can
form a weakly bound state if the overall force between them is repulsive,
since a small dip in the local maximum of the classical potential energy $%
U(a)$ at $a=0$ produces a small barrier around $a=0$, with $a=0$ being a
point of (otherwise) unstable equilibrium, where (excluding the binding
energy effect due to $\Delta\mu$), $U(0)=U_{\text{max}}>0$. Thus, a small
perturbation with energy $E\geq\Delta\mu$ to the bound state can separate
the two kinks at rest. The bound state energy is relatively small since $%
\Delta\mu\propto\beta^{2}$ and $|\beta|\ll\lambda_{\chi},\lambda_{\varphi}$.
The potential energy $U(0)$ of the weakly bound system is then converted
into kinetic energy of the kinks. Of course, more strongly bound states may
exist for $U(0)<0$.

\section{Classical motion}

\ \ \textit{Interaction energy}:\ \ The perturbing potential describing the
interaction between the fields $\chi$ and $\varphi$ is given by (\ref{2})
with $|\beta|\ll\lambda_{\chi}$, $\lambda_{\varphi}$, and we allow $\beta$
to be positive or negative. Since the field corrections $\chi_{1}$ and $%
\varphi_{1}$ are considered to be very small, with the base functions $%
\chi_{0}$ and $\varphi_{0}$ dominating, we now neglect the small corrections
and approximate $V_{1}(\chi,\varphi)$ by $V_{1}(\chi_{0},\varphi_{0})$. The
fields $\chi_{0}$ and $\varphi_{0}$ obey the equations of motion that follow
from $\mathcal{L}_{0}(\chi_{0},\varphi_{0})=\frac{1}{2}(\partial%
\chi_{0})^{2}+\frac{1}{2}(\partial\varphi_{0})^{2}-V_{0}(\chi_{0},%
\varphi_{0})$ with static solutions given by (\ref{e17}). With our notation $%
k_{\chi}=k$, $k_{\varphi }=q$, $x_{\chi}=0$, and $x_{\varphi}=a$, these
solutions take the form%
\begin{equation}
\chi_{0}(x)=\eta\tanh kx,\ \ \ \ \ \varphi_{0}(x)=\sigma\tanh q(x-a)
\label{70}
\end{equation}

The energy density\ (surface energy density for domain walls) associated
with the topological kink\ solutions (\ref{70}) is $T_{00}^{(0)}=-\mathcal{L}%
_{0}(\chi_{0},\varphi_{0})$, and the \textit{residual} energy density $%
\rho_{1}=-\mathcal{L}_{I}(\chi_{0},\varphi_{0})=V_{1}(\chi_{0},\varphi_{0})$
is that associated with the kink\ \textit{interactions}, namely,%
\begin{equation}
\rho_{1}=V_{1}(\chi_{0},\varphi_{0})=\frac{1}{2}\beta\left(
\chi_{0}^{2}-\eta^{2}\right) \left( \varphi_{0}^{2}-\sigma^{2}\right) =\frac{%
1}{2}\beta\eta^{2}\sigma^{2}\text{ sech}^{2}kx\cdot\text{sech}^{2}q(x-a)
\label{71}
\end{equation}

The \textquotedblleft potential energy\textquotedblright\ of interaction $%
U(a)$ (potential energy/unit area, for domain walls) is given by the
integration of $\rho_{1}(x,a)$,%
\begin{equation}
U(a)=\int\rho_{1}(x,a)dx=\frac{1}{2}\beta\eta^{2}\sigma^{2}I(a)=\frac{1}{2}%
\beta\eta^{2}\sigma^{2}\text{ }\int\text{sech}^{2}kx\cdot\text{sech}%
^{2}q(x-a)dx  \label{72}
\end{equation}

This is viewed as the potential energy of the $\varphi_{0}$ kink in the
presence of the $\chi_{0}$ kink \cite{note}.

\bigskip

\ \ The integral $I(a)=\int$sech$^{2}kx\cdot$sech$^{2}q(x-a)dx$ can be
approximated if we take, for example, $k\gtrsim q$, and use (\ref{27}), with
sech$^{2}kx\rightarrow\frac{2}{k}\delta(x)$. Integration gives \cite{note1} $%
I(a)\rightarrow\frac{2}{k}$sech$^{2}qa$. With this approximation we have%
\begin{equation}
U(a)=\frac{1}{2}\beta\eta^{2}\sigma^{2}I(a)=\frac{\beta\eta^{2}\sigma^{2}}{k}%
\ \text{sech}^{2}qa=U_{0}\ \text{sech}^{2}qa,\ \ \ \ \ U_{0}=\frac {%
\beta\eta^{2}\sigma^{2}}{k}  \label{73}
\end{equation}

The sign of the potential energy $U(a)$ is governed by the sign of $\beta$,
so that $U\geq0$ for $\beta>0$ and $U\leq0$ for $\beta<0$. For $\beta>0$ the
position $a=0$ locates a point of unstable equilibrium, while for $\beta<0$
the point $a=0$ is one of stable equilibrium. The maximum magnitude of $U(a)$
is $|U|_{\max}=|\beta|\eta^{2}\sigma^{2}/k$.

\bigskip

\ \ \textit{Interkink force}:\ \ The \textquotedblleft
force\textquotedblright\ of interaction $F_{x}(a)$ (force per unit area for
domain walls) between the two kinks (e.g., the force on $\varphi_{0}$ at $%
x=a $ due to $\chi_{0}$ at $x=0$) is%
\begin{equation}
F_{x}(a)=-\frac{\partial U(a)}{\partial a}=f_{0}\tanh qa\cdot\text{sech}%
^{2}qa,\ \ \ \ \ f_{0}=2qU_{0}=2\beta\eta^{2}\sigma^{2}\frac{q}{k}
\label{74}
\end{equation}

For $\beta>0$ the force is repulsive and for $\beta<0$ the force is
attractive. The magnitude $|F_{x}|$ maximizes at $qa=\frac{1}{2}\ln(2+\sqrt {%
3})\approx\frac{2}{3}$, corresponding to a separation distance between kink
centers of $a\sim\frac{2}{3}w_{\varphi}$, i.e., roughly $2/3\times$ the
width of the $\varphi$ kink.

\bigskip

\ \ \textit{Motion}:\ \ The classical motion of the system then depends upon
whether the potential $U(a)$ is repulsive $(\beta>0)$ or attractive $%
(\beta<0)$, and can then be described as a classical two-body system,
assuming that when dissipative effects due to scalar radiation of the $\chi$
and $\varphi$ fields are neglected, the mechanical energy $E=T(a)+U(a)$ is
conserved, where $T(a)$ is the kinetic energy of the system. Classical
turning points of the two kink system depend upon the total energy $E$ and
the potential energy $|U_{\text{max}}|=|U_{0}|$.

\bigskip

\ \ For a repulsive interaction $(\beta>0)$ turning points can exist for $%
E<U_{0}$, so that the kinks have a minimum distance of approach. We recall,
however, that for the (otherwise) maximum of $U(0)$ there is a small dip at $%
a=0$ due to the $\psi_{1}(x)$ corrections and the associated mass defect $%
\Delta\mu$ of (\ref{69}), which is $O(\beta^{2})$, so that a weakly bound
state can exist even for $\beta>0$.

\bigskip

\ \ On the other hand, for an attractive interaction $(\beta<0)$ the $\chi$
and $\varphi$ kinks can form a bound state. A conserved topological current
density (see, e.g., \cite{Rajaraman},\cite{Vach},\cite{Manton}) is $j^{\mu }=%
\frac{1}{2v}\epsilon^{\mu\nu}\partial_{\nu}\psi$ (with $\psi=\chi$ or $%
\varphi$, $v=\eta$ or $\sigma$, and $\epsilon^{01}=1$), so that the
topological charge is $Q=\frac{1}{v}\int_{-\infty}^{\infty}\partial_{x}\psi
dx=\frac{1}{v}[\psi(\infty)-\psi(-\infty)]=+1$ for $\chi$ or $\varphi$ kinks
and $Q=-1$ for $\bar{\chi}$ or $\bar{\varphi}$ antikinks. The corresponding
charges for $(\chi,\varphi)$ and $(\bar{\chi},\bar{\varphi})$ bound states
are $Q=+2$ and $Q=-2$, respectively, and $Q=0$ for $(\bar{\chi},\varphi)$
and $(\chi,\bar{\varphi})$ states. It should also be pointed out that a
general system containing many kinks and antikinks will accommodate
collisions and annihilations of kinks and antikinks of the same type. The
description of motion in this case is much more complicated. (See, for
example, \cite{Campbell} regarding kink-antikink interactions in the $%
\phi^{4}$ model, and \cite{Alonso1} for kink interactions in a two-component
model).

\section{Meson mixing}

\ The model given by (\ref{1}) and (\ref{2}) has a mass (squared) matrix $%
\boldsymbol{M}^{2}$ given by (\ref{5}) which is associated with the
perturbative \textquotedblleft meson\textquotedblright\ particle
\textquotedblleft flavor\textquotedblright\ states $\chi$ and $\varphi$. For 
$\beta\neq0$ there are off-diagonal terms due to the interacting scalar
fields, indicating that the flavor states are not mass eigenstates. The mass
eigenvalues of $\boldsymbol{M}^{2}$ are given by (\ref{5a}). Let us now
denote $m_{+}=m_{1}$ and $m_{-}=m_{2}$, with $m_{1}>m_{2}$. Then, in terms
of $m_{1,2}$, $(m_{1}>m_{2})$,%
\begin{equation}
\begin{array}{cc}
m_{1}^{2}=\frac{1}{2}\left[ m_{\chi}^{2}+m_{\varphi}^{2}+\sqrt{4\mu
^{4}+(m_{\chi}^{2}-m_{\varphi}^{2})^{2}}\right] \smallskip &  \\ 
m_{2}^{2}=\frac{1}{2}\left[ m_{\chi}^{2}+m_{\varphi}^{2}-\sqrt{4\mu
^{4}+(m_{\chi}^{2}-m_{\varphi}^{2})^{2}}\right] & 
\end{array}
\label{m1}
\end{equation}

and the corresponding mass eigenstates are $\phi_{+}=\phi_{1}$ and $\phi
_{-}=\phi_{2}$, and the flavor states $\chi$ and $\varphi$ are linear
combinations of $\phi_{1}$ and $\phi_{2}$, as shown in (\ref{5b}). Now, for
the individual noninteracting fields $\chi$ and $\varphi$ there are, in
addition to the kink modes, nonperturbative modes including a zero mode $%
\psi_{0}$ and a discrete excitation mode $\psi_{D}$ for each field, along
with the meson radiation modes, labelled here as $\varepsilon_{p}(x,t)$ with
momentum $p$. A meson radiation mode has energy $\omega=\sqrt{p^{2}+m_{\psi
}^{2}}$ with $m_{\psi}^{2}=m_{\chi}^{2},m_{\varphi}^{2}$ and%
\begin{equation}
\varepsilon_{p}(x,t)=f_{p}(x)e^{-i\omega t},\ \ \ \
f_{p}(x)=Ae^{ipx}[3\tanh^{2}z-1-p^{2}w^{2}-iwp\tanh z]  \label{m2}
\end{equation}

with $z=\kappa(x-x_{0})$, $w=1/\kappa$, with $\kappa=k,q$ and $p$ is the
momentum. The asymptotic scattering solutions are given by $f_{p}(x)\propto
e^{ipx}$.

\bigskip

\ \ \textit{Mixing of meson particle states:}\ \ We denote the
(perturbative) meson particle flavor states at time $t=0$ by $%
|\chi(0)\rangle=|\chi\rangle,$ $|\varphi(0)\rangle=|\varphi\rangle$, and the
mass eigenstates at time $t=0$ are $|\phi_{1}(0)\rangle=|\phi_{1}\rangle,\
|\phi_{2}(0)\rangle=|\phi _{2}\rangle$:%
\begin{equation}
\left( 
\begin{array}{c}
|\chi\rangle \\ 
|\varphi\rangle%
\end{array}
\right) =\left( 
\begin{array}{cc}
\cos\theta & \sin\theta \\ 
-\sin\theta & \cos\theta%
\end{array}
\right) \ \left( 
\begin{array}{c}
|\phi_{1}\rangle \\ 
|\phi_{2}\rangle%
\end{array}
\right) =\boldsymbol{R}(\theta)\ \left( 
\begin{array}{c}
|\phi_{1}\rangle \\ 
|\phi_{2}\rangle%
\end{array}
\right)  \label{m3}
\end{equation}

where $\boldsymbol{R}(\theta)$ is the rotation matrix, $\theta$ is the
mixing angle, and the kets $|\phi_{1,2}\rangle$ represent orthonormal states,%
\begin{equation}
\langle\phi_{1}|\phi_{1}\rangle=\langle\phi_{2}|\phi_{2}\rangle =1,\ \ \ \
\langle\phi_{1}|\phi_{2}\rangle=0  \label{m4}
\end{equation}

The energy eigenstates evolve: $|\phi_{i}(t)\rangle=|\phi_{i}(0)\rangle
e^{-iE_{i}t}=|\phi_{i}\rangle e^{-iE_{i}t}$, so that

\begin{equation}
\left( 
\begin{array}{c}
|\phi_{1}(t)\rangle \\ 
|\phi_{2}(t)\rangle%
\end{array}
\right) =\left( 
\begin{array}{c}
|\phi_{1}\rangle e^{-iE_{1}t} \\ 
|\phi_{2}\rangle e^{-iE_{2}t}%
\end{array}
\right) =\boldsymbol{R}^{-1}(\theta)\ \left( 
\begin{array}{c}
|\chi(t)\rangle \\ 
|\varphi(t)\rangle%
\end{array}
\right)  \label{m5}
\end{equation}

and%
\begin{equation}
\left( 
\begin{array}{c}
|\chi(t)\rangle \\ 
|\varphi(t)\rangle%
\end{array}
\right) =\boldsymbol{R}(\theta)\left( 
\begin{array}{c}
|\phi_{1}\rangle e^{-iE_{1}t} \\ 
|\phi_{2}\rangle e^{-iE_{2}t}%
\end{array}
\right) =\left( 
\begin{array}{cc}
\cos\theta & \sin\theta \\ 
-\sin\theta & \cos\theta%
\end{array}
\right) \left( 
\begin{array}{c}
|\phi_{1}\rangle e^{-iE_{1}t} \\ 
|\phi_{2}\rangle e^{-iE_{2}t}%
\end{array}
\right)  \label{m6}
\end{equation}

Therefore,%
\begin{equation}
\begin{array}{ll}
|\chi(t)\rangle & =\cos\theta|\phi_{1}\rangle e^{-iE_{1}t}+\sin\theta|\phi
_{2}\rangle e^{-iE_{2}t} \\ 
|\varphi(t)\rangle & =-\sin\theta|\phi_{1}\rangle e^{-iE_{1}t}+\cos\theta
|\phi_{2}\rangle e^{-iE_{2}t}%
\end{array}
\label{m7}
\end{equation}

These results can be used to write \cite{CMS}%
\begin{equation}
\begin{array}{cc}
|\chi(t)\rangle=(e^{-iE_{1}t}\cos^{2}\theta+e^{-iE_{2}t}\sin^{2}\theta
)|\chi\rangle+\sin\theta\cos\theta(e^{-iE_{2}t}-e^{-iE_{1}t})|\varphi\rangle
&  \\ 
|\varphi(t)\rangle=(e^{-iE_{2}t}-e^{-iE_{1}t})\sin\theta\cos\theta|\chi
\rangle+(e^{-iE_{1}t}\sin^{2}\theta+e^{-iE_{2}t}\cos^{2}\theta)|\varphi%
\rangle & 
\end{array}
\label{m7a}
\end{equation}

\ \ \textit{Probabilities}:\ \ The probability that a $\chi$ meson emitted
at time $t=0$ becomes either a $\chi$ or $\varphi$ meson at time $t$ is \cite%
{CMS},\cite{AC18} 
\begin{subequations}
\label{m8}
\begin{align}
& 
\begin{array}{ll}
P(\chi\rightarrow\chi,t) & =|\langle\chi|\chi(t)\rangle|^{2}=(e^{-iE_{1}t}%
\cos^{2}\theta+e^{-iE_{2}t}\sin^{2}\theta)^{2} \\ 
& =1-\frac{1}{2}\sin^{2}(2\theta)[1-\cos(E_{1}-E_{2})t]%
\end{array}
\label{m8a} \\
& 
\begin{array}{ll}
P(\chi\rightarrow\varphi,t) & =|\langle\varphi|\chi(t)\rangle|^{2}=[\sin
\theta\cos\theta(e^{-iE_{2}t}-e^{-iE_{1}t})]^{2} \\ 
& =\frac{1}{2}\sin^{2}(2\theta)[1-\cos(E_{1}-E_{2})t]%
\end{array}
\label{m8b}
\end{align}

\ \ \textit{High energy limit}:\ \ For ultrarelativistic particles with $%
E\gg m$ and $E\approx E_{1}\approx E_{2}$, write $%
E_{1}-E_{2}=(E_{1}^{2}-E_{2}^{2})/(E_{1}+E_{2})\approx\frac{%
(m_{1}^{2}-m_{2}^{2})}{2E}=\frac{\Delta m^{2}}{2E}$, where $\Delta
m^{2}=m_{1}^{2}-m_{2}^{2}$. Then for an ultrarelativistic particle with
speed $v\approx1$ emitted from $x=0$ at time $t=0$ we have at a distance $x$
a phase for which $(E_{1}-E_{2})t\approx \frac{\Delta m^{2}}{2E}x=\frac{2\pi
x}{L}$, where the oscillation length is $L=4\pi E/\Delta m^{2}$. Therefore,
the probability that an ultrarelativistic $\chi$ particle emitted from $x=0$
at time $t=0$ reaches a stationary $\varphi$ kink located at $x=a$ in the
form of a $\varphi$ particle at time $t$ is 
\end{subequations}
\begin{equation}
P(\chi\rightarrow\varphi,t\approx a)=\frac{1}{2}\sin^{2}(2\theta)\left[
1-\cos\frac{2\pi a}{L}\right]  \label{m9}
\end{equation}

A beam consisting of $N_{\chi}^{(0)}$ ultrarelativistic monoenergetic $\chi$
particles emitted from $x=0$ reaches the $\varphi$ kink at $x=a$ with only a
number of $N_{\chi}(a)\approx N_{\chi}^{(0)}(1-P_{\chi\rightarrow\varphi})$
of $\chi$ particles. The $\varphi$ mesons do not reflect from an unexcited $%
\varphi$ kink (see, e.g.,\cite{Goldstone75,Jackiw77,Rajaraman}), and
therefore do not exert a force upon it. The $\chi$ meson force upon the $%
\varphi$ kink is thus reduced by a factor of $N_{\chi}^{(0)}P_{\chi
\rightarrow\varphi}$, where $P_{\chi\rightarrow\varphi}=P(\chi\rightarrow
\varphi,t\approx a)$.

\bigskip

\ \ \textit{Low energy limit}:\ \ For low energy particles with $%
p_{1}\approx p_{2}\ll m$ and $E\approx m$, the situation is more
complicated, in that it is found that different mass eigenstates which reach
the same position $x=a$ at the same time are actually \textit{emitted} from
the source at \textit{different times} \cite{Kobach18PLB}. This complication
will be further compounded if there is a nontrivial spectrum of energies
associated with the emitted $\chi$ radiation. No attempt, therefore, is made
here to extract any useful quantitative information concerning the actual
force exerted on a $\varphi$ kink by emitted $\chi$ bosons.

\bigskip

\ \textit{Meson-kink interactions}:\ \ Some qualitative remarks may be made,
however, concerning effects of meson-kink interactions. First, if a $\chi$
meson transforms into a $\varphi$ meson when reaching a $\varphi$ kink,
these $\varphi$ mesons do not reflect from the $\varphi$ kink, but merely
experience a phase shift \cite{Goldstone75,Jackiw77,Rajaraman}. Also, high
energy $\chi$ particles with wavelength $\lambda\ll w_{\varphi}$, i.e., $%
p\gg w_{\varphi }^{-1}=q$, have essentially no reflection from the $\varphi$
kink, that is, the reflection coefficient $\mathcal{R}\approx0$ \cite%
{VSbookSec13}. But for very low energy particles with $\lambda\gg
w_{\varphi} $, or $p\ll w_{\varphi }^{-1}=q$, the reflection is strong with $%
\mathcal{R}\approx1$ \cite{VSbookSec13}, so that most very low energy $\chi$
particles are reflected and can therefore produce a scalar radiation force
on the $\varphi$ kink. This force will vary with the probability $%
P_{\chi\rightarrow\varphi}$, which, in turn, will depend upon the position $%
a $ of the $\varphi$ kink.

\section{Summary}

\ \ A Rayleigh-Schr\"{o}dinger perturbation scheme has been developed in
order to study the interactions of kinks or domain walls formed from two
different scalar fields $\chi$ and $\varphi$. This scheme results in
successive sets of corrections to the zero order solitonic solutions $%
\chi_{0}(x)$ and $\varphi_{0}(x)$ satisfying an unperturbed system with
potential $V_{0}(\chi,\varphi)$. The perturbation is introduced through an
interaction potential $V_{1}(\chi,\varphi)$. The particular model studied
here uses the quartic potential $V_{0}=\frac{1}{4}\lambda_{\chi}(\chi^{2}-%
\eta^{2})^{2}+\frac{1}{4}\lambda_{\varphi}(\varphi^{2}-\sigma^{2})^{2}$ and
an interaction potential $V_{1}=\frac{1}{2}\beta(\chi^{2}-\eta^{2})(\varphi
^{2}-\sigma^{2})$. The unperturbed static solutions $\chi_{0}(x)$ and $%
\varphi_{0}(x)$ are represented by the usual $Z_{2}$ kinks. The first order
corrections $\chi_{1}(x)$ and $\varphi_{1}(x)$ are found, which exhibit the
peculiar property that the interaction induces each kink to form a
condensate within the other kink. Therefore the $\chi$ kink acquires a $%
\varphi$ condensate, and the $\varphi$ kink acquires a $\chi$ condensate.

\bigskip

\ \ The masses of these condensates are determined, and it is reasoned that
the condensate masses decrease with separation distance between the kinks,
and vanishes when the kinks coincide. The associated mass defect implies the
possible existence of a weakly bound state when the overall interkink force
is repulsive. When subjected to a small disturbance, the weakly bound
two-kink state can fission into two separate kinks with kinetic energies.

\bigskip

\ \ A classical potential energy of the system and an interkink force are
defined, allowing a qualitative description of the classical motion of the
system. The interkink force can be either attractive or repulsive, depending
upon the sign of the coupling control parameter $\beta$. The system can
therefore accommodate scattering states and bound states. In the case of an
attractive interaction force, the bound states can be much more tightly
bound than the weakly bound states associated with a repulsive potential,
with the composite two-kink bound state having topological charge of $Q=\pm2$
or $0$.

\bigskip

\ \ Finally, it is pointed out that an interaction between the $\chi$ and $%
\varphi$ scalar fields generally results in a nondiagonal mass matrix,
indicating that the $\chi$ and $\varphi$ \textquotedblleft
flavor\textquotedblright\ states are actually linear combinations of mass
eigenstates $\phi_{1}$ and $\phi_{2}$ of the fields. As a consequence, there
are oscillations of the flavor states as the mesons from the position $x=0$
of the $\chi$ kink propagate to the position $x=a$ of the $\varphi$ kink.
Time-dependent probabilities $P(\chi\rightarrow\chi,t)$ and $P(\chi
\rightarrow\varphi,t)$ are found for a $\chi$ boson to be found as a $\chi$
or a $\varphi$ boson at time $t$. For ultrarelativistic particles a standard
result is given for the probabilities and oscillation lengths. However, the
situation is much murkier for the case of nonrelativistic particles. At any
rate, the $\chi$ radiation force exerted upon the $\varphi$ kink will be
reduced by an amount that depends upon the meson mixing probabilities.

\appendix

\section{Perturbation expansion scheme}

\ \ As written in (\ref{9})-(\ref{11}), we can expand $\psi(x,g)$ in powers
of $g$, and expand $H(\psi)$ about the unperturbed (base) solution $\psi_{0}$%
:

\ \ 
\begin{equation}
\begin{array}{ll}
\psi(x,g) & =\psi_{0}(x)+\delta\psi(x,g) \\ 
\delta\psi(x,g) & =\sum_{n=1}^{\infty}g^{n}\psi_{n}(x)=g\psi_{1}(x)+g^{2}%
\psi_{2}(x)+\cdot\cdot\cdot%
\end{array}
\label{A1}
\end{equation}

with $\psi=\chi,\varphi$, and%
\begin{equation}
H(\chi,\varphi)=H(\psi_{0})+\left( \delta\chi\partial_{\chi}+\delta
\varphi\partial_{\varphi}\right) H(\chi,\varphi)\Big|_{\chi_{0},\varphi_{0}}+%
\frac{1}{2}\left( \delta\chi\partial_{\chi}+\delta\varphi\partial_{\varphi
}\right) ^{2}H(\chi,\varphi)\Big|_{\chi_{0},\varphi_{0}}+\cdot\cdot \cdot
\label{A2}
\end{equation}

where $H=F,G$, and $(\delta\psi\partial_{\psi})^{2}H=(\delta\psi)^{2}%
\partial_{\psi}^{2}H$ and so on. Since $H=H_{0}+gH_{1}$ this becomes%
\begin{align}
H(\chi,\varphi) & =H_{0}(\psi_{0})+\left(
\delta\chi\partial_{\chi}+\delta\varphi\partial_{\varphi}\right)
H_{0}(\chi,\varphi)\Big|_{\chi _{0},\varphi_{0}}+\frac{1}{2}\left(
\delta\chi\partial_{\chi}+\delta \varphi\partial_{\varphi}\right)
^{2}H_{0}(\chi,\varphi)\Big|_{\chi _{0},\varphi_{0}}+\cdot\cdot\cdot  \notag
\\
& +gH_{1}(\psi_{0})+g\left( \delta\chi\partial_{\chi}+\delta\varphi
\partial_{\varphi}\right) H_{1}(\chi,\varphi)\Big|_{\chi_{0},\varphi_{0}}+%
\frac{1}{2}g\left( \delta\chi\partial_{\chi}+\delta\varphi\partial
_{\varphi}\right) ^{2}H_{1}(\chi,\varphi)\Big|_{\chi_{0},\varphi_{0}}+\cdot%
\cdot\cdot  \label{A3}
\end{align}

\bigskip

\ \ The equations of motion for the full system (\ref{3}) can now be written
in expanded form \ with the aid of (\ref{A1})-(\ref{A3}):%
\begin{equation}
\begin{array}{ll}
\square(\chi_{0}+g\chi_{1}+g^{2}\chi_{2}+g^{3}\chi_{3}+\cdot\cdot\cdot
)+F_{0}(\chi_{0},\varphi_{0})+gF_{1}(\chi_{0},\varphi_{0}) &  \\ 
+\left[ \left( g\chi_{1}+g^{2}\chi_{2}+g^{3}\chi_{3}+\cdot\cdot\cdot\right)
\partial_{\chi}F_{0}(\chi_{0},\varphi_{0})\right] +\left[ \left(
g\varphi_{1}+g^{2}\varphi_{2}+g^{3}\varphi_{3}+\cdot\cdot\cdot\right)
\partial_{\varphi}F_{0}(\chi_{0},\varphi_{0})\right] &  \\ 
+\frac{1}{2}\left[ \left( g\chi_{1}+g^{2}\chi_{2}+\cdot\cdot\cdot\right)
^{2}\partial_{\chi}^{2}F_{0}(\chi_{0},\varphi_{0})+\left(
g\varphi_{1}+g^{2}\varphi_{2}+\cdot\cdot\cdot\right)
^{2}\partial_{\varphi}^{2}F_{0}(\chi_{0},\varphi_{0})\right] &  \\ 
+\left[ \left( g\chi_{1}+g^{2}\chi_{2}+\cdot\cdot\cdot\right) \left(
g\varphi_{1}+g^{2}\varphi_{2}+\cdot\cdot\cdot\right)
\partial_{\chi}\partial_{\varphi}F_{0}(\chi_{0},\varphi_{0})\right] &  \\ 
+g\left[ \left( g\chi_{1}+g^{2}\chi_{2}+g^{3}\chi_{3}+\cdot\cdot
\cdot\right) \partial_{\chi}F_{1}(\chi_{0},\varphi_{0})\right] +g\left[
\left( g\varphi_{1}+g^{2}\varphi_{2}+g^{3}\varphi_{3}+\cdot\cdot\cdot\right)
\partial_{\varphi}F_{1}(\chi_{0},\varphi_{0})\right] &  \\ 
+\frac{1}{2}g\left[ \left( g\chi_{1}+g^{2}\chi_{2}+\cdot\cdot\cdot\right)
^{2}\partial_{\chi}^{2}F_{1}(\chi_{0},\varphi_{0})+\left(
g\varphi_{1}+g^{2}\varphi_{2}+\cdot\cdot\cdot\right)
^{2}\partial_{\varphi}^{2}F_{1}(\chi_{0},\varphi_{0})\right] &  \\ 
+g\left[ \left( g\chi_{1}+g^{2}\chi_{2}+\cdot\cdot\cdot\right) \left(
g\varphi_{1}+g^{2}\varphi_{2}+\cdot\cdot\cdot\right)
\partial_{\chi}\partial_{\varphi}F_{1}(\chi_{0},\varphi_{0})\right] =0 & 
\end{array}
\label{A4}
\end{equation}

and similarly for the $\varphi$ equation of motion with $\square
\chi\rightarrow\square\varphi$ and $F(\chi,\varphi)\rightarrow G(\chi
,\varphi)$. The various $g^{n}$ terms can be collected to give the equations
for the $\chi_{n}$ and $\varphi_{n}$.

\begin{equation}
\begin{array}{ll}
g^{0}: & \square\chi_{0}+F_{0}(\chi_{0},\varphi_{0})=0 \\ 
g^{1}: & \square\chi_{1}+(\chi_{1}\partial_{\chi}+\varphi_{1}\partial
_{\varphi})F_{0}(\chi_{0},\varphi_{0})+F_{1}(\chi_{0},\varphi_{0})=0 \\ 
g^{2}: & \square\chi_{2}+\left(
\chi_{2}\partial_{\chi}+\varphi_{2}\partial_{\varphi}\right)
F_{0}(\chi_{0},\varphi_{0})+\frac{1}{2}\left(
\chi_{1}^{2}\partial_{\chi}^{2}+\varphi_{1}^{2}\partial_{\varphi}^{2}\right)
F_{0}(\chi_{0},\varphi_{0}) \\ 
& +\chi_{1}\varphi_{1}\partial_{\chi}\partial_{\varphi}F_{0}(\chi_{0},%
\varphi_{0})+\left( \chi_{1}\partial_{\chi}+\varphi_{1}\partial_{\varphi
}\right) F_{1}(\chi_{0},\varphi_{0})=0%
\end{array}
\label{A5}
\end{equation}

\begin{equation}
\begin{array}{ll}
g^{0}: & \square\varphi_{0}+G_{0}(\chi_{0},\varphi_{0})=0 \\ 
g^{1}: & \square\varphi_{1}+(\chi_{1}\partial_{\chi}+\varphi_{1}\partial_{%
\varphi})G_{0}(\chi_{0},\varphi_{0})+G_{1}(\chi_{0},\varphi_{0})=0 \\ 
g^{2}: & \square\varphi_{2}+\left( \chi_{2}\partial_{\chi}+\varphi
_{2}\partial_{\varphi}\right) G_{0}(\chi_{0},\varphi_{0})+\frac{1}{2}\left(
\chi_{1}^{2}\partial_{\chi}^{2}+\varphi_{1}^{2}\partial_{\varphi}^{2}\right)
G_{0}(\chi_{0},\varphi_{0}) \\ 
& +\chi_{1}\varphi_{1}\partial_{\chi}\partial_{\varphi}G_{0}(\chi_{0},%
\varphi_{0})+\left( \chi_{1}\partial_{\chi}+\varphi_{1}\partial_{\varphi
}\right) G_{1}(\chi_{0},\varphi_{0})=0%
\end{array}
\label{A6}
\end{equation}

For our model given by (\ref{1}) and (\ref{2}), the first order equations
for $\psi_{1}(x)$ are given by (\ref{12}), with $H_{0}(\chi,\varphi)$ and $%
H_{1}(\chi,\varphi)$ given by (\ref{15}).

\section{Approximate first order corrections}

\ \ \textbf{Solution for }$\boldsymbol{\chi}_{1}$\textbf{:} \ We have the
approximate second order nonhomogeneous differential equation (DE)%
\begin{equation}
\psi^{\prime\prime}(x)-4k^{2}\psi(x)+12k\delta(x)\psi(x)=-\frac{2B_{1}}{q}%
\tanh kx\cdot\delta(x-a)  \label{B1}
\end{equation}

with the solutions%
\begin{equation}
\begin{array}{ll}
\text{I:\ \ \ }\psi_{1}=Ae^{2kx}, & x<0 \\ 
\text{II: \ }\psi_{2}=Be^{2kx}+Ce^{-2kx}, & 0<x<a \\ 
\text{III: }\psi_{3}=De^{-2kx}, & x>a%
\end{array}
\label{B2}
\end{equation}

\ \ The continuity of $\psi$ at $x=0$ and $x=a$ gives the following
constraints:%
\begin{equation}
\begin{array}{ll}
\psi_{1}(0)=\psi_{2}(0): & A=B+C \\ 
\psi_{2}(a)=\psi_{3}(a): & Be^{2ka}+Ce^{-2ka}=De^{-2ka}%
\end{array}
\label{B3}
\end{equation}

Now upon integrating the DE (\ref{B1}) about $x=\pm\epsilon$ (where the
right hand side is absent) and about $x=a\pm\epsilon$ (where the $%
12k\delta(x)$ term is absent) and taking the limit $\epsilon\rightarrow0$,
we obtain%
\begin{equation}
\begin{array}{ll}
x\approx0: & \int_{-\epsilon}^{\epsilon}\psi^{\prime\prime}(x)dx-4k^{2}%
\int_{-\epsilon}^{\epsilon}\psi(x)dx+12k\int_{-\epsilon}^{\epsilon}\psi(x)%
\delta(x)dx=0 \\ 
x\approx a: & \int_{a-\epsilon}^{a+\epsilon}\psi^{\prime\prime}(x)dx-4k^{2}%
\int_{a-\epsilon}^{a+\epsilon}\psi(x)dx=-\frac{2B_{1}}{q}\int_{a-\epsilon
}^{a+\epsilon}\tanh kx\delta(x-a)dx%
\end{array}
\label{B4}
\end{equation}

Keeping in mind that $\psi(x)$ is continuous, so that as $\epsilon
\rightarrow0$, $\int_{-\epsilon}^{\epsilon}\psi(x)dx=0$ and $%
\int_{a-\epsilon }^{a+\epsilon}\psi(x)dx=0$, we have%
\begin{equation}
\begin{array}{ll}
\psi_{2}^{\prime}(0)-\psi_{1}^{\prime}(0)+12k\psi_{1}(0)=0 & \implies
B-C+5A=0 \\ 
\psi_{3}^{\prime}(a)-\psi_{2}^{\prime}(a)=-\frac{2B_{1}}{q}\tanh ka & 
\implies De^{-2ka}+Be^{2ka}-Ce^{-2ka}=\frac{B_{1}}{kq}\tanh ka%
\end{array}
\label{B5}
\end{equation}

The four constraint equations given by (\ref{B3}) and (\ref{B5}) allow us to
determine%
\begin{equation}
A=-\frac{1}{2}B,\ \ \ C=-\frac{3}{2}B,\ \ \ D=\left( e^{4ka}-\frac{3}{2}%
\right) B,\ \ \ B=\frac{B_{1}}{2kq}e^{-2ka}\tanh ka  \label{B6}
\end{equation}

From (\ref{e17}) we have $k_{\chi}=k=\sqrt{\lambda_{\chi}/2}\cdot\eta$ and $%
k_{\varphi}=q=\sqrt{\lambda_{\varphi}/2}\cdot\sigma$. This in conjunction
with $B_{1}=\beta\eta\sigma^{2}$ allows us to write the coefficient $B$ as%
\begin{equation}
B=\frac{B_{1}}{2kq}e^{-2ka}\tanh ka=\frac{\beta\eta\sigma^{2}}{\sqrt {%
\lambda_{\chi}\lambda_{\varphi}}\eta\sigma}e^{-2ka}\tanh ka=\frac{\beta
\sigma}{\sqrt{\lambda_{\chi}\lambda_{\varphi}}}e^{-2ka}\tanh ka  \label{B7}
\end{equation}

Then (\ref{B2}), (\ref{B6}), and (\ref{B7}) give us%
\begin{equation}
\chi_{1}(x)\approx\frac{\beta\sigma}{\sqrt{\lambda_{\chi}\lambda_{\varphi}}}%
e^{-2ka}\tanh ka\times\left\{ 
\begin{array}{ll}
-\frac{1}{2}e^{2kx}, & x<0 \\ 
e^{2kx}-\frac{3}{2}e^{-2kx}, & 0<x<a \\ 
\left( e^{4ka}-\frac{3}{2}\right) e^{-2kx}, & x>a%
\end{array}
\right\}  \label{B8}
\end{equation}

The correction $|\chi_{1}|$ maximizes at $x=a$, the location of the $\varphi$
kink (FIG. 1). For $x=a$ the correction is, approximately,%
\begin{equation}
\chi_{1}(a)\approx\frac{\beta\sigma\tanh ka}{\sqrt{\lambda_{\chi}\lambda_{%
\varphi}}}(1-\tfrac{3}{2}e^{-4ka})\sim\frac{\beta\sigma}{\sqrt{%
\lambda_{\chi}\lambda_{\varphi}}}  \label{B9}
\end{equation}

for $ka\gtrsim1$ ($a\gtrsim w_{\chi}$). The requirement that $|\chi_{1}(a)|$
is dominated by $|\chi_{0}(a)|\sim\eta$ for $ka\gtrsim1$, i.e., $|\chi
_{1}|/\eta\ll1$, then translates into the requirement $\left( |\beta |/\sqrt{%
\lambda_{\chi}\lambda_{\varphi}}\right) \sigma/\eta\ll1$. For $%
\sigma/\eta\sim O(1)$ this means that the approximate solution is valid
provided that%
\begin{equation}
\frac{|\beta|}{\sqrt{\lambda_{\chi}\lambda_{\varphi}}}\frac{\sigma}{\eta}\sim%
\frac{|\beta|}{\sqrt{\lambda_{\chi}\lambda_{\varphi}}}\ll 1,\ \ \ \text{for}%
\ \frac{\sigma}{\eta}\sim O(1)  \label{B10}
\end{equation}

which is in accord with the original assumption that $|\beta|\ll\lambda_{%
\chi }$, $\lambda_{\varphi}$, so that the perturbing potential $V_{1}$ is a
small perturbation to the unperturbed potential $V_{0}$.\ \ 

\bigskip

\ \ \textbf{Solution for }$\boldsymbol{\varphi}_{1}$\textbf{:}\ \ Now denote 
$\varphi_{1}$ by $\varphi_{1}(x)=\psi(x)$, and again $k_{\chi}=k$, $%
k_{\varphi}=q$, $x_{\chi}=0$, and $x_{\varphi}=a$. The location of the $\chi$
kink is $x=0$, and that of the $\varphi$ kink is $x=a$, as before. Also
define the constant $B_{2}=\beta\eta^{2}\sigma$. Using the same
approximations as before, we divide the $x$ space into three regions with
functions $\psi _{1}(x)$, $\psi_{2}(x)$, and $\psi_{3}(x)$ in regions I, II,
and III, respectively. With the delta function approximation, (\ref{24b}) is
written as%
\begin{equation}
\psi^{\prime\prime}(x)-4q^{2}\psi(x)+12q\delta(x-a)\psi(x)=-\frac{2B_{2}}{k}%
\tanh q(x-a)\delta(x)  \label{B11}
\end{equation}

with boundary conditions $\psi\rightarrow0$ as $x\rightarrow\pm\infty$. Each
region is again $\delta$ function-free, and the solutions are again of
exponential form $e^{\pm2qx}$. Specifically,%
\begin{equation}
\begin{array}{ll}
\text{I:\ \ \ }\psi_{1}=Ae^{2qx}, & x<0 \\ 
\text{II: \ }\psi_{2}=Be^{2qx}+Ce^{-2qx}, & 0<x<a \\ 
\text{III: }\psi_{3}=De^{-2qx}, & x>a%
\end{array}
\label{B12}
\end{equation}

where the coefficients $A,B,C,D$ are now new ones for the $\varphi_{1}$
function. We use continuity of $\psi(x)$ at $x=0$ and $x=a,$ and integrate
the DE (\ref{B11}) to obtain constraints on $\psi^{\prime}(0)$ and $%
\psi^{\prime }(a)$.

\bigskip

\ \ The continuity of $\psi$ at $x=0$ and $x=a$ gives the following
constraints:%
\begin{equation}
\begin{array}{ll}
\psi_{1}(0)=\psi_{2}(0): & A=B+C \\ 
\psi_{2}(a)=\psi_{3}(a): & Be^{2qa}+Ce^{-2qa}=De^{-2qa}%
\end{array}
\label{B13}
\end{equation}

Integration of the DE (\ref{B11}) about $x=\pm\epsilon$ and about $%
x=a\pm\epsilon$ and taking the limit $\epsilon\rightarrow0$, we obtain%
\begin{equation}
\begin{array}{ll}
\psi_{2}^{\prime}(0)-\psi_{1}^{\prime}(0)=\frac{2B_{2}}{k}\tanh qa & 
\implies B-C-A=\frac{B_{2}}{kq}\tanh qa \\ 
\psi_{3}^{\prime}(a)-\psi_{2}^{\prime}(a)+12q\psi_{3}(a)=0 & \implies
5De^{-2qa}=Be^{2qa}-Ce^{-2qa}%
\end{array}
\label{B14}
\end{equation}

The constraint equations given by (\ref{B13}) and (\ref{B14}) yield%
\begin{equation}
A=\left( 1-\frac{3}{2}e^{-4qa}\right) C,\ \ \ B=-\frac{3}{2}Ce^{-4qa},\ \ \
D=-\frac{1}{2}C,\ \ \ C=-\frac{B_{2}}{2kq}\tanh qa  \label{B15}
\end{equation}

Using $k=\sqrt{\lambda_{\chi}/2}\cdot\eta$ and $q=\sqrt{\lambda_{\varphi}/2}%
\cdot\sigma$, along with $B_{2}=\beta\eta^{2}\sigma$ allows us to write the
coefficient $C$ as%
\begin{equation}
C=-\frac{B_{2}}{2kq}\tanh qa=-\frac{\beta\eta^{2}\sigma}{\sqrt{\lambda_{\chi
}\lambda_{\varphi}}\eta\sigma}\tanh qa=-\frac{\beta\eta}{\sqrt{\lambda_{\chi
}\lambda_{\varphi}}}\tanh qa  \label{B16}
\end{equation}

Then (\ref{B12}), (\ref{B15}), and (\ref{B16}) give%
\begin{equation}
\varphi_{1}(x)\approx-\frac{\beta\eta}{\sqrt{\lambda_{\chi}\lambda_{\varphi}}%
}\tanh qa\times\left\{ 
\begin{array}{ll}
\left( 1-\frac{3}{2}e^{-4qa}\right) e^{2qx}, & x<0 \\ 
-\frac{3}{2}e^{-4qa}e^{2qx}+e^{-2qx}, & 0<x<a \\ 
-\frac{1}{2}e^{-2qx}, & x>a%
\end{array}
\right\}  \label{B17}
\end{equation}

The correction $|\varphi_{1}|$ maximizes at $x=0$, the location of the $\chi$
kink (FIG. 2). The assumption that $|\varphi_{1}|\ll\sigma$ is satisfied if $%
(|\beta|/\sqrt{\lambda_{\chi}\lambda_{\varphi}})\eta/\sigma\ll1$. So the
approximate solution for $\varphi_{1}$ is valid provided that%
\begin{equation}
\frac{|\beta|}{\sqrt{\lambda_{\chi}\lambda_{\varphi}}}\frac{\eta}{\sigma}\sim%
\frac{|\beta|}{\sqrt{\lambda_{\chi}\lambda_{\varphi}}}\ll 1,\ \ \ \text{for}%
\ \frac{\eta}{\sigma}\sim O(1)  \label{B18}
\end{equation}

Therefore (\ref{B10}) and (\ref{B18}) imply that the approximate solutions
for $\chi_{1}$ and $\varphi_{1}$ are valid if $|\beta|\ll\sqrt{\lambda_{\chi
}\lambda_{\varphi}}$ and $\sigma\sim\eta$, which again is in accord with the
original assumption that $|\beta|\ll\lambda_{\chi}$, $\lambda_{\varphi}$, so
that the perturbing potential $V_{1}$ is a small perturbation to the
unperturbed potential $V_{0}$.

\bigskip

\ \ Once again, the particular solution of (\ref{B17}) is accompanied by the
zero mode solution $\varphi_{1}^{(0)}(x)\propto\varphi_{0}^{\prime}(x)\sim$
sech$^{2}q(x-a)$ which solves the homogeneous (sourceless) DE of\ (\ref{24b}%
), but since it has nothing to do with the $\chi-\varphi$ interaction we
dismiss it from further consideration.

\section{Condensate masses}

\ \ \textbf{\textquotedblleft Mass\textquotedblright\ of the }$\boldsymbol{%
\chi}_{1}$ \textbf{condensate:}\ \ The energy-momentum tensor for the $\chi$
field is%
\begin{equation}
T_{\mu\nu}^{\chi}=\partial_{\mu}\chi\partial_{\nu}\chi-\eta_{\mu\nu }%
\mathcal{L}_{\chi}  \label{49}
\end{equation}

where $\mathcal{L}_{\chi}=\mathcal{L}_{\chi}^{(0)}+\mathcal{L}_{I}$, 
\begin{equation}
\mathcal{L}_{\chi}^{(0)}=\frac{1}{2}(\partial\chi)^{2}-\frac{1}{4}%
\lambda_{\chi}(\chi^{2}-\eta^{2})^{2},\ \ \ \ \ \mathcal{L}_{I}=-V_{1}=-%
\frac{1}{2}\beta(\chi^{2}-\eta^{2})(\varphi^{2}-\sigma^{2})  \label{50}
\end{equation}

The idea is to calculate the energy-momentum $T_{\mu\nu}^{\chi}$ that arises
from the interaction of the $\chi$ and $\varphi$ kinks. This means that we
dismiss any contributions that arise from the pure zero modes $%
\chi_{1}^{(0)}(x)$ and $\varphi_{1}^{(0)}(x)$, as these modes are solutions
of the homogeneous (i.e., sourceless) DEs for $\chi_{1}$ and $\varphi_{1}$,
and therefore do not arise from the $\chi-\varphi$ interaction.

\bigskip

\ \ For the region near the $\varphi_{0}$ kink, $x=a$, we take $\chi(x\sim
a)\approx\eta+\chi_{1}(x)$. (We can also note from (\ref{47}) that near $%
x\sim a$ there is a tiny peak in $\varphi_{1}$ with $\varphi_{1}(x\sim
a)\propto\beta\eta e^{-2qa}$ which we neglect for $qa\gtrsim1,$ so that $%
\varphi_{1}(x\sim a)\approx0$.) Keeping in mind that $|\chi_{1}|\ll\eta$ and
retaining dominant terms results in 
\begin{equation}
T_{00}^{\chi}(x\sim a)=-\mathcal{L}_{\chi}(x\sim a)\approx\frac{1}{2}(\chi
_{1}^{\prime})^{2}+\lambda_{\chi}\eta^{2}\chi_{1}^{2}+\beta\eta\sigma^{2}%
\chi_{1}\text{ sech}^{2}q(x-a)  \label{51}
\end{equation}

where $^{\prime}=\partial_{x}=\partial/\partial x$. This $%
T_{00}^{\chi}(x\sim a)$ is then the energy density associated with the $%
\chi_{1}$ field, which is concentrated near $x=a$. An integration of this
energy density then gives the mass $\Sigma_{\chi}$ of the $\chi_{1}$
condensate. Referring back to (\ref{38}) for the solution of $\chi_{1}$, we
note that for $ka\gtrsim1$ (or $a\gtrsim w_{\chi}$), that $%
e^{-2ka}e^{2kx}\ll1$ for $x<0$, and the solution for $x<0$ can be ignored.
Furthermore, for $ka\gtrsim1$ and $x\sim a$ we have $e^{-2kx}\ll e^{2kx}$
and $e^{4ka}\gg\frac{3}{2}$ so that (\ref{38}) simplifies to%
\begin{equation}
\chi_{1}(x\sim a)\approx B_{0}e^{-2ka}\times\left\{ 
\begin{array}{ll}
e^{2kx}, & x<a \\ 
e^{4ka}e^{-2kx}, & x>a%
\end{array}
\right\}  \label{52}
\end{equation}

where%
\begin{equation}
B_{0}=\frac{\beta\sigma}{\sqrt{\lambda_{\chi}\lambda_{\varphi}}}\tanh ka
\label{53}
\end{equation}

Using (\ref{52}) and (\ref{53}) to evaluate (\ref{51}) for $x\sim a$ leads to%
\begin{equation}
T_{00}^{\chi}\approx4k^{2}B_{0}^{2}\times\left\{ 
\begin{array}{l}
e^{-4ka}e^{4kx} \\ 
e^{4ka}e^{-4kx}%
\end{array}
\right\} +\beta\eta\sigma^{2}B_{0}\text{ sech}^{2}q(x-a)\times\left\{ 
\begin{array}{l}
e^{-2ka}e^{2kx} \\ 
e^{2ka}e^{-2kx}%
\end{array}
\right\} ,\ \ \ \left\{ 
\begin{array}{c}
x<a \\ 
x>a%
\end{array}
\right\}  \label{54}
\end{equation}

A cumbersome integral can be avoided by again using the delta function
approximation (\ref{27}), sech$^{2}q(x-a)\rightarrow\frac{2}{q}\delta(x-a)$
with $\frac{1}{q}=\sqrt{\frac{2}{\lambda_{\varphi}}}\frac{1}{\sigma}$. We
now have%
\begin{equation}
T_{00}^{\chi}\approx4k^{2}B_{0}^{2}\times\left\{ 
\begin{array}{l}
e^{-4ka}e^{4kx} \\ 
e^{4ka}e^{-4kx}%
\end{array}
\right\} +\frac{2}{q}\beta\eta\sigma^{2}B_{0}\text{ }\delta(x-a)\times
\left\{ 
\begin{array}{l}
e^{-2ka}e^{2kx} \\ 
e^{2ka}e^{-2kx}%
\end{array}
\right\} ,\ \ \ \left\{ 
\begin{array}{c}
x<a \\ 
x>a%
\end{array}
\right\}  \label{55}
\end{equation}

This can now be integrated to obtain $\Sigma _{\chi }=\int_{-\infty
}^{a}T_{00}^{\chi }dx+\int_{a}^{\infty }T_{00}^{\chi }dx$. The integrand
appearing with the delta function has a value of $1$ for $x\rightarrow a$
and is continuous at $x=a$, so that $\int_{-\infty }^{\infty }e^{\pm
2kx}e^{\mp 2ka}\delta (x-a)dx=\int_{-\infty }^{\infty }e^{-2k|x-a|}\delta
(x-a)dx\rightarrow 1$. Therefore%
\begin{align}
\Sigma _{\chi }& \approx 4k^{2}B_{0}^{2}\left( \int_{-\infty
}^{a}e^{-4ka}e^{4kx}dx+\int_{a}^{\infty }e^{4ka}e^{-4kx}dx\right) +\frac{2}{q%
}\beta \eta \sigma ^{2}B_{0}  \notag \\
& =2kB_{0}^{2}+\frac{2}{q}\beta \eta \sigma ^{2}B_{0}  \label{56}
\end{align}

Using (\ref{53}), along with $k=\sqrt{\frac{\lambda_{\chi}}{2}}\eta$, $q=%
\sqrt{\frac{\lambda_{\varphi}}{2}}\sigma$, the approximate mass of the $%
\chi_{1}$ condensate (\ref{56}) is%
\begin{equation}
\Sigma_{\chi}(a)\approx2\beta^{2}\eta\sigma^{2}\left[ \sqrt{\frac {%
\lambda_{\chi}}{2}}\frac{\tanh^{2}ka}{\lambda_{\chi}\lambda_{\varphi}}+\sqrt{%
\frac{2}{\lambda_{\varphi}}}\frac{\tanh ka}{\sqrt{\lambda_{\chi}\lambda_{%
\varphi}}}\right]  \label{57}
\end{equation}

Although we have assumed, for ease of computation, that $ka\gtrsim1$, we
might reasonably extrapolate to the case $ka<1$ or $ka\rightarrow0$. In that
case we find that $\Sigma_{\chi}$ decreases and approaches zero when the $%
\chi$ and $\varphi$ kinks overlap with their centers coinciding.

\bigskip

\ \ \textbf{\textquotedblleft Mass\textquotedblright\ of the }$\boldsymbol{%
\varphi}_{1}$\textbf{\ condensate:}\ \ We follow the same procedure to
obtain the \textquotedblleft mass\textquotedblright\ (surface energy for a
domain wall) $\Sigma_{\varphi}$ of the $\varphi_{1}$ condensate. Again, we
dismiss any contributions from the pure zero modes $\chi_{1}^{(0)}$ and $%
\varphi_{1}^{(0)}$, as these do not arise from the $\chi-\varphi$
interaction. We write the energy-momentum tensor%
\begin{equation}
T_{\mu\nu}^{\varphi}=\partial_{\mu}\varphi\partial_{\nu}\varphi-\eta_{\mu\nu
}\mathcal{L}_{\varphi}  \label{58}
\end{equation}

where $\mathcal{L}_{\varphi}=\mathcal{L}_{\varphi}^{(0)}+\mathcal{L}_{I}$,%
\begin{equation}
\mathcal{L}_{\varphi}^{(0)}=\frac{1}{2}(\partial\varphi)^{2}-\frac{1}{4}%
\lambda_{\varphi}(\varphi^{2}-\sigma^{2})^{2},\ \ \ \ \ \mathcal{L}%
_{I}=-V_{1}=-\frac{1}{2}\beta(\chi^{2}-\eta^{2})(\varphi^{2}-\sigma^{2})
\label{59}
\end{equation}

For the region near the $\chi_{0}$ kink, $x=0$, we take $\varphi
(x\sim0)\approx-\sigma+\varphi_{1}(x)$. Again, $|\varphi_{1}|\ll\sigma$, and
retaining dominant terms gives 
\begin{equation}
T_{00}^{\varphi}(x\sim0)=-\mathcal{L}_{\varphi}(x\sim0)\approx\frac{1}{2}%
(\varphi^{\prime})^{2}+\lambda_{\varphi}\sigma^{2}\varphi_{1}^{2}-\beta
\eta^{2}\sigma\varphi_{1}\text{ sech}^{2}kx  \label{60}
\end{equation}

To obtain the mass $\Sigma_{\varphi}$ we integrate the energy density $%
T_{00}^{\varphi}(x\sim0)$ associated with the $\varphi_{1}$ condensate
residing within the $\chi_{0}$ host kink. We can use (\ref{47}) to examine $%
\varphi_{1}(x\sim0)$. In the neighborhood of $x\sim0$, with $ka\gtrsim1$ and 
$qa\gtrsim1$, we write, approximately,%
\begin{equation}
\varphi_{1}(x\sim0)\approx C\times\left\{ 
\begin{array}{cc}
e^{2qx}, & x<0 \\ 
e^{-2qx}, & x>0%
\end{array}
\right\}  \label{61}
\end{equation}

where%
\begin{equation}
C=-\frac{\beta\eta}{\sqrt{\lambda_{\chi}\lambda_{\varphi}}}\tanh qa
\label{62}
\end{equation}

From (\ref{60})-(\ref{62}) we get, for $x\sim0$,%
\begin{equation}
T_{00}^{\varphi}\approx4q^{2}C^{2}\times\left\{ 
\begin{array}{c}
e^{4qx} \\ 
e^{-4qx}%
\end{array}
\right\} -\beta\eta^{2}\sigma C\text{ sech}^{2}kx\times\left\{ 
\begin{array}{c}
e^{2qx} \\ 
e^{-2qx}%
\end{array}
\right\} ,\ \ \ \left\{ 
\begin{array}{c}
x<0 \\ 
x>0%
\end{array}
\right\}  \label{63}
\end{equation}

where use has been made of $\lambda_{\varphi}\sigma^{2}=2q^{2}$ in the first
term. We again use the delta function approximation sech$^{2}kx\rightarrow 
\frac{2}{k}\delta(x)=2\sqrt{\frac{2}{\lambda_{\chi}}}\frac{1}{\eta}\delta(x)$%
:%
\begin{equation}
T_{00}^{\varphi}\approx4q^{2}C^{2}\times\left\{ 
\begin{array}{c}
e^{4qx} \\ 
e^{-4qx}%
\end{array}
\right\} -2\sqrt{\frac{2}{\lambda_{\chi}}}\beta\eta\sigma C\delta
(x)\times\left\{ 
\begin{array}{c}
e^{2qx} \\ 
e^{-2qx}%
\end{array}
\right\} ,\ \ \ \left\{ 
\begin{array}{c}
x<0 \\ 
x>0%
\end{array}
\right\}  \label{64}
\end{equation}

Integration then gives%
\begin{align}
\Sigma _{\varphi }& \approx 4q^{2}C^{2}\left( \int_{-\infty
}^{0}e^{4qx}dx+\int_{0}^{\infty }e^{-4qx}dx\right) -2\sqrt{\frac{2}{\lambda
_{\chi }}}\beta \eta \sigma C\int_{-\infty }^{\infty }e^{-2q|x|}\delta (x)dx
\notag \\
& =2qC^{2}-2\sqrt{\frac{2}{\lambda _{\chi }}}\beta \eta \sigma C  \label{65}
\end{align}

Using $q=\sqrt{\frac{\lambda_{\varphi}}{2}}\sigma$ along with (\ref{62}), (%
\ref{65}) can be written as%
\begin{equation}
\Sigma_{\varphi}(a)\approx2\beta^{2}\eta^{2}\sigma\left[ \sqrt{\frac {%
\lambda_{\varphi}}{2}}\frac{\tanh^{2}qa}{\lambda_{\chi}\lambda_{\varphi}}+%
\sqrt{\frac{2}{\lambda_{\chi}}}\frac{\tanh qa}{\sqrt{\lambda_{\chi}\lambda_{%
\varphi}}}\right]  \label{66}
\end{equation}

\ \ The results for the \textquotedblleft masses\textquotedblright\ $\Sigma
_{\chi }$ and $\Sigma _{\varphi }$, given by (\ref{57}) and (\ref{66}) again
allow us to reasonably expect that each mass decreases with decreasing
separation distance $a$, presumably to zero when the centers of the two
kinks coincide. Such a decrease in the total system mass suggests the
presence of a very weak ($\propto \beta ^{2}$), but nonzero, force of
attraction between the two kinks, allowing a weakly bound state to possibly
exist. This attractive force must be of fairly short range, since $\tanh
\kappa a$ approaches unity for $\kappa a\sim 2$, where $\kappa =k=1/w_{\chi }
$ or $\kappa =q=1/w_{\varphi }$.

\bigskip 

\textbf{Acknowledgement:} I wish to thank an anonymous referee for useful
comments.

\ \

\bigskip


\begin{thebibliography}{99}
\bibitem{note1} Actually, we take the approximation to be sufficient for $%
\frac{k}{q}\gtrsim2$.
\end{thebibliography}
\end{document}